\documentclass{emulateapj-rtx4}
\usepackage{natbib}
\usepackage{epsf,graphics}
\usepackage{subfigure,graphicx}
\usepackage{bm}
\usepackage{amssymb}
\usepackage{amsbsy}
\usepackage{amsmath}
\usepackage{gensymb}

\begin{document}

\title{A robust mass estimator for dark matter subhalo perturbations in strong 
gravitational lenses}

\author{Quinn E. Minor}
\affiliation{Department of Science, Borough of Manhattan Community College, 
City University of New York, New York, NY 10007, USA}
\affiliation{Department of Astrophysics, American Museum of Natural History, 
New York, NY 10024, USA}
\author{Manoj Kaplinghat}
\affiliation{Department of Physics and Astronomy, University of California, 
Irvine CA 92697, USA}
\author{Nan Li}
\affiliation{Department of Astronomy and Astrophysics, The University of Chicago, 5640 South Ellis Avenue, Chicago, IL 60637, USA}
\affiliation{High Energy Physics Division, Argonne National Laboratory, Lemont,  IL 60439, USA}

\begin{abstract}
A few dark matter substructures have recently been detected in strong 
gravitational lenses though their perturbations of highly magnified images.  We 
derive a characteristic scale for lensing perturbations and show that this is 
significantly larger than the perturber's Einstein radius. We show that the 
perturber's projected mass enclosed within this radius, scaled by the log-slope 
of the host galaxy's density profile, can be robustly inferred even if the 
inferred density profile and tidal radius of the perturber are biased.  We 
demonstrate the validity of our analytic derivation by using several 
gravitational lens simulations where the tidal radii and the inner log-slopes 
of the density profile of the perturbing subhalo are allowed to vary.  By 
modeling these simulated data we find that our mass estimator, which we call 
the effective subhalo lensing mass, is accurate to within about 10\% or smaller 
in each case, whereas the inferred total subhalo mass can potentially be biased 
by nearly an order of magnitude.  We therefore recommend that the effective 
subhalo lensing mass be reported in future lensing reconstructions, as this 
will allow for a more accurate comparison with the results of dark matter 
simulations.

\end{abstract}

\keywords{gravitational lensing: strong -- dark matter -- galaxies: dwarf\vspace{1.0mm}}

\section{Introduction}\label{sec:intro}

A robust prediction of the standard Cold Dark Matter (CDM) paradigm is that 
galaxy-sized dark matter halos are populated by thousands of bound subhalos 
left over from the hierarchical assembly process 
\citep{diemand2008,navarro2010,springel2008}.  As is well known, far fewer 
satellite galaxies of the Milky Way have been observed compared to the number 
seen in CDM simulations (the so-called ``missing satellites problem''; 
\citealt{klypin1999}).  Additionally, the most massive observed Milky Way 
satellite galaxies seem to have significantly low central densities compared to 
massive satellites in CDM simulations; this is the ``too-big-to-fail problem'' 
\citep{boylan2011,boylan2012}, and the related ``core-cusp problem'' 
\citep{flores1994,moore1994,kuzio2006}.  These problems may be entirely 
resolved in the context of CDM through baryonic feedback effects, such as 
supernova feedback \citep{governato2012} or stellar quenching through 
photoevaporation during reionization \citep{bullock2000}.  However, it is also 
possible that the solution is partially due to a modification of the standard 
collisionless CDM paradigm.  For example, self-interacting dark matter (SIDM) 
has been invoked to explain the too-big-to-fail problem through the formation 
of constant-density cores at the centers of subhalos 
\citep{rocha2013,elbert2014,spergel2000}.  Warm dark matter (WDM) has been 
proposed to explain the missing satellites problem, in which the thermal motion 
of free-streaming dark matter particles erase small-scale structure 
\citep{lovell2014,bode2001,abazajian2006}.  However, because of uncertainties 
in initial conditions and the star formation physics involved, it is difficult 
to robustly estimate the effect that stellar feedback should have on the 
subhalos of observed Milky Way satellite galaxies.  Thus, constraining physical 
properties of the dark matter particle using observed local group satellites is 
a daunting task.

Strong gravitational lensing is an alternative approach that shows great 
promise in being able to constrain the properties of dark matter substructure 
at high redshift 
\citep{mao1998,metcalf2001,dalal2002,moustakas2003,kochanek2004,keeton2009,koopmans2005,vegetti2009,nierenberg2014,xu2015,cyrracine2016}.  
Since even completely dark subhalos can be lensed, lensing offers the promise 
of observing substructure where little or no star formation has occurred. This 
allows us to constrain not only the abundance and mass function of subhalos 
that might otherwise be invisible \citep{vegetti2014b}, but also (in principle) 
the density profile of subhalos for whom stellar feedback effects may be 
relatively small \citep{vegetti2014}.  

In the past several years, a few detections of dark substructures in 
gravitational lenses have already been reported by observing their effect on 
highly magnified images.  Two of these were discovered in the SLACS dataset 
\citep{vegetti2010,vegetti2012}, while more recently, a subhalo was reported by 
\cite{hezaveh2016} to have been detected in an ALMA image of the lens system 
SDP.81 \citep{alma2015}.  Comparing the masses reported in these studies to 
subhalo properties in cosmological simulations, however, is not 
straightforward.  Nearly all of the parametric subhalo modeling studies to date 
(\citealt{vegetti2010}, \citealt{vegetti2012}, \citealt{hezaveh2016}, 
\citealt{suyu2010}) have modeled the subhalo as a smoothly truncated singular 
isothermal sphere using the so-called Pseudo-Jaffe profile (the one exception 
we are aware of being \citealt{nierenberg2014}); however, since subhalos are 
typically quite dark matter-dominated, there is no compelling theoretical 
expectation for them to have an isothermal profile. CDM simulations suggest 
that the subhalos are more likely to have a Navarro-Frenk-White (NFW) profile 
\citep{navarro1996} or Einasto profile.  In addition, the assumed tidal radius 
of the subhalo may be incorrect. For example, if the true tidal radius is 
greater than the assumed tidal radius, there would be unaccounted-for mass 
further out in the lens plane. It would also affect the shape of the projected 
density profile of the subhalo even at smaller radii, because there would be a 
differing amount of mass along the line of sight.  Thus, since the assumed 
density profile is unlikely to perfectly match the subhalo's true density 
profile, it is important to establish whether the total subhalo mass can be 
determined robustly; and if not, whether there is a characteristic scale within 
which the enclosed mass \emph{is} robust to changes in the subhalo's density 
profile.

In this article, we show that the subhalo masses thus far reported in 
gravitational lens systems may be significantly biased, mainly due to 
unrealistic assumptions about the subhalos' tidal radius (which is also pointed 
out by \citealt{vegetti2014b}), but also due to a mismatched subhalo density 
profile.  By fitting to simulated lens data, we will show that there exists a 
characteristic scale for subhalo lensing, which we call the subhalo 
perturbation radius, within which the subhalo's estimated mass \emph{is} 
approximately invariant to changes in the subhalo's density profile and tidal 
radius.

We introduce the subhalo perturbation radius in Section 
\ref{sec:subhalo_perturbation}, and discuss its general properties, including 
the robust subhalo mass estimator defined by this radius, which call the 
effective subhalo lensing mass.  In Section \ref{sec:tidal_radius} we discuss 
the assumptions underlying the subhalo's tidal radius in gravitational lens 
models, and compare these to theoretical expectations for dark matter subhalos.  
In Section \ref{sec:application} we describe the gravitational lens simulations 
we use to demonstrate the effective subhalo lensing mass. We present the 
results of our analysis in Section \ref{sec:results}, and show that the
effective subhalo lensing mass is well-reproduced in each case. In Section 
\ref{sec:range} we explore the range of validity of the effective lensing mass 
for different subhalo masses and positions. Finally, in Section 
\ref{sec:discussion} we discuss our recommendations for modeling subhalos to 
obtain reliable mass estimates, and conclude with our main points in Section 
\ref{sec:conclusions}.  Analytic formulas for the subhalo perturbation radius, 
along with the effective subhalo lensing mass and corresponding derivations, 
are given in the appendices.

\section{A characteristic scale for subhalo lensing perturbations} 
\label{sec:subhalo_perturbation}

\subsection{The size of subhalo perturbations and defining a subhalo's 
effective lensing mass}

The perturbation in a lensed image produced by a subhalo is only evident if 
that subhalo lies near the tangential critical curve of the lens, distorting 
the critical curve. Near the warped critical curve, the surface brightness is 
altered and additional images may even appear. The size of this critical curve 
distortion is comparable to the scale at which the lensed surface brightness is 
significantly perturbed by the subhalo.

We define the \emph{subhalo perturbation radius} to be the distance from the 
subhalo to the critical curve along the direction where the magnification is 
perturbed the most by the subhalo. This radius, which we denote as 
$\delta\theta_{c}$, can be loosely thought of as the ``radius of maximum 
warping'' of the critical curve, i.e. the furthest distance by which the 
subhalo ``pushes'' the critical curve outward compared to the position of the 
critical curve without the subhalo present. As we show in Appendix 
\ref{sec:appendix_a}, $\delta\theta_{c}$ can be found by solving the following 
equation for $r$:

\begin{equation}
1 - \kappa_0(\boldsymbol{\theta_s}+\boldsymbol{r}) - \Gamma_{tot}(\boldsymbol{\theta_s}+\boldsymbol{r}) = \bar\kappa_s(r)
\label{kapav_eq}
\end{equation}

Here, $\kappa_0(\boldsymbol\theta)$ refers to the projected density profile of 
the host galaxy, $\Gamma_{tot}$ is the magnitude of the total unperturbed 
shear, i.e.  the complex sum of the shear from the host galaxy 
$\Gamma_0(\boldsymbol\theta)$ plus the external shear $\Gamma_{ext}$ created by 
neighboring galaxies, and $\bar\kappa_s(r)$ is the average projected density of 
the subhalo a distance $r$ from the center of the subhalo. The direction of 
$\boldsymbol{r}$ corresponds to the direction along which the subhalo's shear 
aligns with that of the total unperturbed shear $\boldsymbol{\Gamma_{tot}}$. As 
long as the external shear is not too large ($\Gamma_{ext} \lesssim 0.2$), then 
to good approximation $\boldsymbol{r}$ simply lies along the radial direction 
with respect to the host galaxy (this is discussed in detail in Appendix 
\ref{sec:appendix_b}1).

Since the projected subhalo mass enclosed within a given radius is simply 
$m_{sub}(r) = \bar\kappa(r)\pi r^2 \Sigma_{cr}$, it is evident that the 
solution $\delta\theta_{c}$ to Eq.~\ref{kapav_eq} depends on the subhalo mass 
enclosed within $r=\delta\theta_{c}$, regardless of how the subhalo mass is 
distributed within that radius. Thus, it follows that if a subhalo is detected 
in a gravitational lens and if the best-fit model reproduces both the host halo 
parameters and the subhalo perturbation radius $\delta\theta_{c}$ well, then 
the inferred subhalo mass within $\delta\theta_{c}$ will be accurate 
\emph{regardless of the actual density profile of the subhalo}.

In reality, unless the subhalo density profile perfectly matches the model, the 
host halo parameters will also be perturbed in order to better fit the lensed 
images in the vicinity of the subhalo. To get an idea of how this affects the 
inferred subhalo mass, let us first assume an idealized scenario where the host 
galaxy is axisymmetric and no external shear is present. In this case, 
Eq.~\ref{kapav_eq} can be written as

\begin{equation}
1 - \bar\kappa_0(\theta_s + r) = \bar\kappa_s(r),
\label{kapav_eq_axial}
\end{equation}
where we have used the fact that for an axisymmetric lens, $\kappa_0 + \Gamma_0 
= \bar\kappa_0$. Note the direction of maximum warping is simply the radial 
direction in this case. If we assume the host galaxy's projected density 
follows a power-law profile, then we can write

\begin{equation}
\kappa_0(\theta) = \frac{2-\alpha}{2}\left(\frac{b}{\theta}\right)^\alpha
\end{equation}
where $b$ is the Einstein radius of the host galaxy. We then have
\begin{equation}
\bar\kappa_0(\theta) = \left(\frac{b}{\theta}\right)^\alpha.
\end{equation}

In the vicinity of the subhalo, the host galaxy's deflection profile 
$\bar\kappa_0 \theta$ can be Taylor expanded to first order, which gives (after 
dividing by $\theta = \theta_s + r$)

\begin{equation}
\bar\kappa_0(\theta_s + r) \approx \left(\frac{b}{\theta_s}\right)^\alpha \left[1 - \alpha\frac{r}{\theta_s+r}\right].
\label{kappa0_axial_eq}
\end{equation}
Note that in the isothermal case ($\alpha=1$), this formula becomes exact; this 
is because the deflection is constant for an isothermal profile, 	so 
higher-order terms vanish anyway in that case (indeed, this is our motivation 
for expanding the deflection instead of expanding $\bar\kappa_0$ directly).

Let us assume the subhalo is close to the critical curve, but not exactly on 
it; more precisely, if the distance between the subhalo and the unperturbed 
critical curve is $\Delta\theta \equiv b - \theta_s$, then we assume that 
$\Delta\theta/b \ll 1$. (If the subhalo is not close enough to the critical 
curve to satisfy this condition, its effects are unlikely to be observable in 
any case!) If we substitute $b = \theta_s + \Delta\theta$ into 
Eq.~\ref{kappa0_axial_eq}, expand to first order in $\Delta\theta/b$ and 
substitute the result into Eq.~\ref{kapav_eq_axial}, we find that

\begin{equation}
\alpha\left(-\frac{\Delta\theta}{b} + \frac{r}{\theta_s+r}\right) \approx \bar\kappa_s(r),
\end{equation}
and hence
\begin{equation}
1 - \bar\kappa_{0,iso}(\theta_s+r) \approx \frac{\bar\kappa_s(r)}{\alpha}
\end{equation}
where $\kappa_{0,iso}$ refers to the isothermal ($\alpha=1$) profile. This 
equation tells us that if the host galaxy deviates from isothermal, then we 
will get the same subhalo perturbation radius if we have an isothermal host and 
the same subhalo mass scaled by $1/\alpha$. For the more general and realistic 
case where the host galaxy has some ellipticity and external shear is present, 
we show in Appendix \ref{sec:appendix_b}2 that the inferred subhalo mass 
$m_{sub}(\delta\theta_c)$ still scales as $1/\alpha$, with only a small amount 
of error due to the external shear (see Eq.~\ref{invariant_mass_eq} and 
following discussion). Thus, when fitting to lens data, as long as our model 
reproduces the position of the subhalo well, we can say that

\begin{equation}
\frac{m_{sub,fit}(\delta\theta_{c})}{\alpha_{fit}} \approx \frac{m_{sub,true}(\delta\theta_{c})}{\alpha_{true}}.
\end{equation}

We therefore define the \emph{effective subhalo lensing mass} $\tilde m_{sub} 
\equiv m_{sub}(\delta\theta_{c})/\alpha$. We will demonstrate by fitting to 
simulated data in Section \ref{sec:results} that, unlike the total subhalo 
mass, $\tilde m_{sub}$ can indeed be inferred robustly even if the subhalo 
density profile is modeled inaccurately.

\subsection{Subhalo perturbation radius for an isothermal subhalo}

It is useful to get an idea of the scale of $\delta\theta_{c}$. If both the 
host galaxy and the subhalo are modeled as a singular isothermal spheres, then 
Eq.~\ref{kapav_eq_axial} becomes

\begin{equation}
1 - \frac{b}{\theta_s+r} = \frac{b_s}{r},
\label{isosphere_eq}
\end{equation}
where $b$ is the Einstein radius of the host galaxy and $b_s$ is the Einstein 
radius of the subhalo. This leads to a quadratic equation in $r$; if we assume 
that the subhalo is right on the critical curve (i.e., $\theta_s = b$) and that 
$b_s \ll b$, the solution is approximately

\begin{equation}
\delta\theta_{c} \approx \sqrt{b b_s} + \frac{1}{2}b_s,
\label{rmax_approx_eq}
\end{equation}
where $b$ is the Einstein radius of the host galaxy and $b_s$ is the Einstein 
radius of the subhalo. Although the subhalo's projected mass may extend well 
beyond this radius, typically it is only in the neighborhood of 
$\delta\theta_{c}$ that the subhalo perturbation will be apparent and 
distinguishable from a smooth lens model.

Eq.~\ref{rmax_approx_eq} is only accurate in the very idealized case we have 
described. In general, the subhalo will not sit exactly on the unperturbed 
critical curve, and the host galaxy will have some ellipticity as well as 
external shear from neighboring galaxies, all of which will tend to reduce the 
magnification and hence make $\delta\theta_{c}$ somewhat smaller. A general 
analytic formula is derived in Appendix \ref{sec:appendix_b} that takes all of 
these effects into account (see Eq.~\ref{rmax_general}).  For those who are not 
in the lens modeling business, however, it is useful to have a simpler 
approximate formula which is accurate to within 30\% in realistic lens 
scenarios. We recommend the following formula, (which is derived in Appendix 
\ref{sec:appendix_b}):

\begin{equation}
\delta\theta_{c} \approx \sqrt{\frac{\epsilon\theta_s b_s}{\alpha}}\left(1 - 
\frac{x}{2} + \frac{x^2}{8}\right) + \frac{\epsilon b_s}{2\alpha}(1-x)
\label{rmax_approx_eq2}
\end{equation}
where
\begin{equation}
x = \xi\sqrt{\frac{\theta_s}{b\epsilon\alpha}}.
\label{x_eq}
\end{equation}
In this formula, $\theta_s$ is the projected angular separation between the 
subhalo and the center of the host galaxy, and $\alpha$ is the log-slope of the 
host galaxy's density profile. The constants $\epsilon$ and $\xi$ are 
determined by the tidal radius of the subhalo. If the subhalo's distance to the 
host galaxy's center $r_0$ (including the component along the line of sight) is 
much larger than the Einstein radius $b$, then we have $\epsilon \approx 1$, 
$\xi \approx 0$. On the other hand, if we make the assumption that $r_0 \approx 
b$, as is often assumed in lens modeling studies, then we have $\epsilon = 
0.879$, $\xi = 0.293$.  Otherwise, for a given assumed value for $r_0$ (or 
equivalently $r_t$), $\epsilon$ and $\xi$ can be calculated using 
Eqs.~\ref{epsilon_eq}, \ref{xi_eq}, and \ref{beta_eq}.

We find that Eq.~\ref{rmax_approx_eq2} is accurate to within 1\% if the subhalo 
sits exactly on the (unperturbed) critical curve, provided the external shear 
$\Gamma_{ext} \lesssim 0.2$. The further away the subhalo is from the critical 
curve, the worse this approximation becomes---generally it will give a value 
for $\delta\theta_c$ that is too large. In the simulated lenses we examine in 
Section \ref{sec:results}, Eq.~\ref{rmax_approx_eq2} is greater than the true 
$\delta\theta_c$ by about 17-25\% depending on the simulation; by comparison, 
the more precise Eq.~\ref{rmax_general} is accurate to within 2\% (see Table 
\ref{tab:dtheta_c_comparison} in Appendix \ref{sec:appendix_b} for a comparison 
between the different estimates of $\delta\theta_c$). However, in the simulated 
lenses we examine in Section \ref{sec:results}, we will find that the optimal 
radius where the effective subhalo lensing mass can be robustly estimated is a 
little bit larger than the actual value for $\delta\theta_c$ in each case, and 
this may tend to be generally the case. One can see from Figure 
\ref{fig:mprofiles} that Eq.~\ref{rmax_approx_eq2} actually works better than 
using the actual $\delta\theta_c$. Therefore, we generally recommend using this 
approximation and calculating the effective lensing mass within this radius.  
Lens modelers who desire a more precise value can solve Eq.~\ref{kapav_eq} 
numerically or use the analytic formula Eq.~\ref{rmax_general}.

\section{Expectations for the subhalo tidal radius}\label{sec:tidal_radius}

In the majority of subhalo modeling studies that have been performed to date, 
the subhalo's tidal radius is assumed to be given by the formula $r_t \approx 
\sqrt{b b_s}$, where $b$ is the Einstein radius of the host galaxy and $b_s$ is 
the Einstein radius of the subhalo \citep{metcalf2001}. This formula is derived 
from the formula for the Jacobi radius of an orbiting satellite, under the 
assumption that the subhalo has an isothermal density profile and the subhalo's 
distance to the center of the host galaxy is roughly comparable to the host 
galaxy's Einstein radius $b$. The latter assumption is motivated by the fact 
that any subhalo that significantly perturbs the observed surface brightness of 
the galaxy being lensed is likely to lie near the critical curve, and hence its 
projected distance to the center of the host galaxy is approximately equal to 
the Einstein radius.  Thus, if we assume the line-of-sight component of its 
displacement from the host galaxy's center is of the same order-of-magnitude as 
the component within the lens plane, its distance is roughly equal to $b$ (or 
$b\sqrt{2}$ in some studies).

Before we discuss whether this approximation is well-justified, it is worth 
noting that this formula for the tidal radius ($r_t = \sqrt{b b_s}$) is similar 
in scale to the subhalo perturbation radius (Eq.~\ref{rmax_approx_eq}). Thus, 
if the tidal radius is accurately described by this formula, then the majority 
of the subhalo's mass would lie within the subhalo perturbation radius. This 
would be extremely fortunate: since the subhalo's projected mass within the 
subhalo perturbation radius can be robustly inferred (up to the scaling 
$1/\alpha$), it would follow that the total inferred subhalo mass would also be 
fairly robust, regardless of the assumed density profile of the subhalo---in 
other words, the subhalo's total mass and its effective lensing mass would be 
approximately the same. We would then be able to conclude that the total 
subhalo mass estimated in previous gravitational lens studies should have 
little bias due to any mismatch between the subhalo's density profile in the 
model compared to reality.

Unfortunately however, the assumption that $r_0 \approx b$, i.e. that the 
subhalo's distance to the host galaxy is approximately equal to the Einstein 
radius, is likely to be inaccurate based on the results of $\Lambda$CDM 
simulations.  To take an example, in \cite{vegetti2012}, the inferred Einstein 
radius of the host galaxy is $b\approx 0.45$ arcseconds, which at its redshift 
$z_L=0.881$ translates to $\approx 3.6$ kpc.  In dark matter-only CDM 
simulations, only a minute fraction of subhalos exist this close to the center 
of the host galaxy, because they are severely tidally disrupted at this radius.  
For example, in Figure 11 of \cite{springel2008}, virtually no subhalos exist 
closer than 30 kpc to the host halo, regardless of subhalo mass, and the vast 
majority are $\sim$ 100 kpc away or further.

In reality, the presence of baryons make subhalos less susceptible to tidal 
stripping by steepening their inner profiles, and dynamical friction with the 
stellar population of the subhalos may increase the chance of bringing subhalos 
closer in towards the host halo center. However, \cite{despali2016} analyze 
hydrodynamical simulations that include baryons and find a similar radial 
distribution of subhalos compared to dark matter-only simulations, but with 
\emph{fewer} subhalos at small radii (see Figure 3 of their article).  
Observational results using Sloan Digital Sky Survey (SDSS) data have produced 
conflicting results; on the one hand, some studies 
\citep{nierenberg2012,guo2012,chen2009} show that satellite galaxies do appear 
to trace the dark matter density profile and thus can be found relatively close 
to the galaxy center, while others \citep{wojtak2013,budzynski2012,hansen2005} 
suggest a shallower number density profile.  More recently, \cite{wang2014} 
find that for primary galaxies with stellar mass $M_*>10^{11} M_\odot$, the 
number density profile of satellites is shallower than the dark matter profile.  
In any case, such surveys tend to focus on relatively bright satellites, for 
which stellar feedback plays a bigger role.  For the dark subhalos that are 
being detected in gravitational lens systems, it is more likely that the 
subhalo is at a considerably greater distance from the host galaxy compared to 
the Einstein radius, and hence the tidal radius is larger than assumed in the 
$r_0=b$ model. 

\begin{table*}[t]
\centering
\begin{tabular}{|l|c|c|c|l|}
\hline
Lens Scenario & $\gamma$ & $r_0$ (kpc) & $r_t$ (kpc) & Description of subhalo \\
\hline
Simulation 1 & 2 & 7.4 & 1.04 & Isothermal (inner slope -2), distance to host galaxy = Einstein radius ($b$=7.4 kpc) \\
Simulation 2 & 1.5 & 7.4 & 1.13 & Inner slope (-1.5) shallower than isothermal, distance to host galaxy = Einstein radius \\
Simulation 3 & 1 & 7.4 & 1.26 & Inner slope (-1) shallower than isothermal, distance to host galaxy = Einstein radius \\
Simulation 4 & 2 & 100 & 12.0 & Isothermal, distance to host galaxy is 100 kpc \\
\hline
\end{tabular}
\caption{Description of each simulated gravitational lens to be modeled in 
Section \ref{sec:results}. In each simulation, the subhalo's inner density 
profile is of the form $\rho(r) \propto r^{-\gamma}$, and the tidal truncation 
radius $r_t$ is determined by the subhalo's distance $r_0$ to the host galaxy's 
center (Eq.~\ref{rt_eq_corecusp}). When we fit a lens model to each scenario, 
we model the subhalo with a tidally truncated isothermal ($\gamma=2$) profile, 
where the distance to the host galaxy is assumed to be equal to the Einstein 
radius ($b$ = 7.4 kpc). Thus, compared to the model, Simulation 1 has the same 
subhalo model, Simulations 2-3 have a shallower density profile, and Simulation 
4 has a much larger tidal radius. Otherwise, the position of the subhalo, the 
host galaxy parameters, and the external shear are all the same in each lens 
simulation.}
\label{tab:sims}
\end{table*}

\begin{figure}
	\centering
	\includegraphics[height=0.8\hsize,width=1.0\hsize]{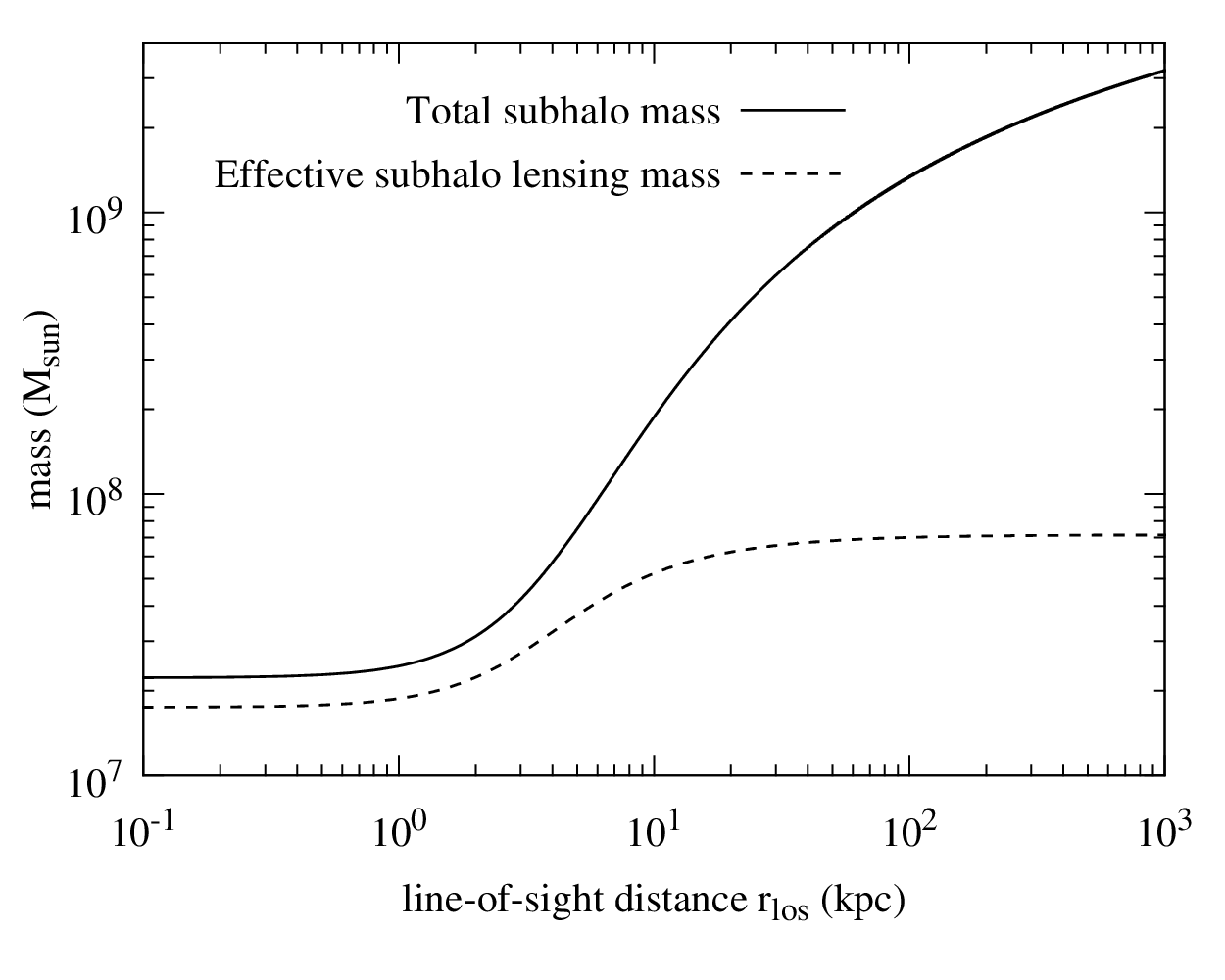}
\caption{Estimated subhalo mass from the NFW solution in \cite{vegetti2014} as 
a function of the subhalo's line-of-sight distance to the lens plane $r_{los}$, 
plotted as the total subhalo mass (solid line) and the effective subhalo 
lensing mass (dashed line), defined as the projected mass of the subhalo 
enclosed within the subhalo perturbation radius.  Note that at large $r_{los}$, 
where the subhalo has suffered relatively little tidal stripping, the total 
subhalo mass can be greater than the effective lensing mass by more than an 
order of magnitude.}
\label{fig:m_rlos}
\end{figure}

To get an idea of the degree of tidal stripping an NFW subhalo would endure at 
this distance, consider an NFW subhalo which is truncated at its tidal radius.  
\cite{vegetti2014} find that the subhalo is consistent with having a $v_{max} 
\approx 28.5$ km/s and $r_{max} \approx 3.1$ kpc. Assuming an NFW subhalo, we 
can use this to get an approximate tidal radius $r_t$ from the formula for the 
Jacobi radius.  If we define $x = r_t/r_s$, the following equation results,

\begin{equation}
\alpha x^3 = \log(1+x) - \frac{x}{1+x},
\label{rt_eq}
\end{equation}
where
\begin{eqnarray}
\alpha \approx & 9.8\times 
10^{-4}\left(\frac{10^{-9}\Sigma_{cr}}{\mathrm{M_\odot kpc}^{-1}}\right)\left(\frac{v_{max}}{10~\mathrm{km/s}}\right)^{-2} \left(\frac{r_{max}}{\mathrm{kpc}}\right)^{2} & \nonumber \\ & \times \left(\frac{b}{\mathrm{arcsec}}\right)\left(\frac{r_0}{100 \mathrm{kpc}}\right)^{-2}\left(\frac{D_L}{\textrm{Gpc}}\right)^{-1}, &
\end{eqnarray}
and $r_0$ is the assumed distance to the host halo. In this case we are setting 
$r_0 \approx b \approx 3.6$ kpc. When this equation is solved numerically, we 
find the tidal radius is $r_t \approx 0.34$ kpc, which is much smaller than 
$r_{max}$. Such extreme tidal stripping would imply that roughly 99\% of the 
subhalo's original virial mass has been stripped away, and almost certainly 
would result in the disruption of the subhalo. Indeed, in Figure 15 of 
\cite{springel2008}, there are no subhalos with a tidal radius smaller than 1 
kpc, at any mass. By comparison, if a subhalo with the same parameters is 
placed at $r_0$ = 100 kpc from the host halo, solving Eq.~\ref{rt_eq} gives a 
tidal radius $r_t \approx 12.6$ kpc, which is far more reasonable.

To get an idea of how much the total subhalo mass can differ compared to the 
effective lensing mass, in Figure \ref{fig:m_rlos} we plot both these 
quantities as a function of the subhalo's line-of-sight distance to the lens 
plane, $r_{los} \approx \sqrt{r_0^2 - b^2}$ for the aforementioned NFW solution 
in \cite{vegetti2014}. The total subhalo mass at each $r_{los}$ is found by 
solving Eq.~\ref{rt_eq} for the tidal radius, then calculating the total mass 
using the smoothly truncated NFW profile from \cite{baltz2009}, while the 
subhalo perturbation radius $\delta\theta_c$ is estimated using 
Eq.~\ref{rmax_approx_eq}. At small $r_{los}$, the tidal radius is only somewhat 
larger than the subhalo perturbation radius, and hence the total mass of the 
subhalo is somewhat larger than the effective lensing mass.  It follows that 
for $r_{los} \lesssim b\approx 3.6$ kpc, the total subhalo mass is relatively 
robust.  As $r_{los}$ increases beyond $b$, the tidal radius becomes 
significantly larger than the subhalo perturbation radius and the effective 
lensing mass ``freezes out''; for $r_{los} > $30 kpc, the subhalo's total mass 
is at least an order of magnitude greater than the effective lensing mass. In 
this regime, which is a much more realistic subhalo distance, any estimate of 
the total mass of the subhalo is likely to be significantly biased due to lack 
of knowledge of the true $r_{los}$.

There is some good news, however: even if the assumption $r_0 \approx b$ is 
inaccurate, it is still the case that if the subhalo is modeled using this 
assumption, then the total inferred subhalo mass should nevertheless be of the 
same order-of-magnitude as the effective lensing mass, which \emph{is} robust.  
In other words, while the subhalo's true total mass may be different from the 
best-fit model's total subhalo mass, the latter may still be interpreted as an 
order-of-magnitude estimate of the projected mass of the subhalo enclosed 
within the subhalo perturbation radius $\delta\theta_c$. This may explain why a 
non-parametric reconstruction of the subhalo potential in \cite{vegetti2012}, 
which did \emph{not} assume a particular density profile for the subhalo, gave 
an estimated subhalo mass which was of a similar magnitude as the total mass 
estimated using the parametric $r_0 = b$ model. This can be expected to hold, 
even though the total subhalo mass may in fact be considerably larger than this 
value. In Section \ref{sec:results}, we will demonstrate this by modeling a 
simulated gravitational lens with a subhalo 100 kpc from the host galaxy using 
the $r_0=b$ assumption.

\section{Application to simulated gravitational lenses}\label{sec:application}

\begin{figure*}[t]
	\centering
	\subfigure[source plane]
	{
		\includegraphics[height=0.37\hsize,width=0.48\hsize]{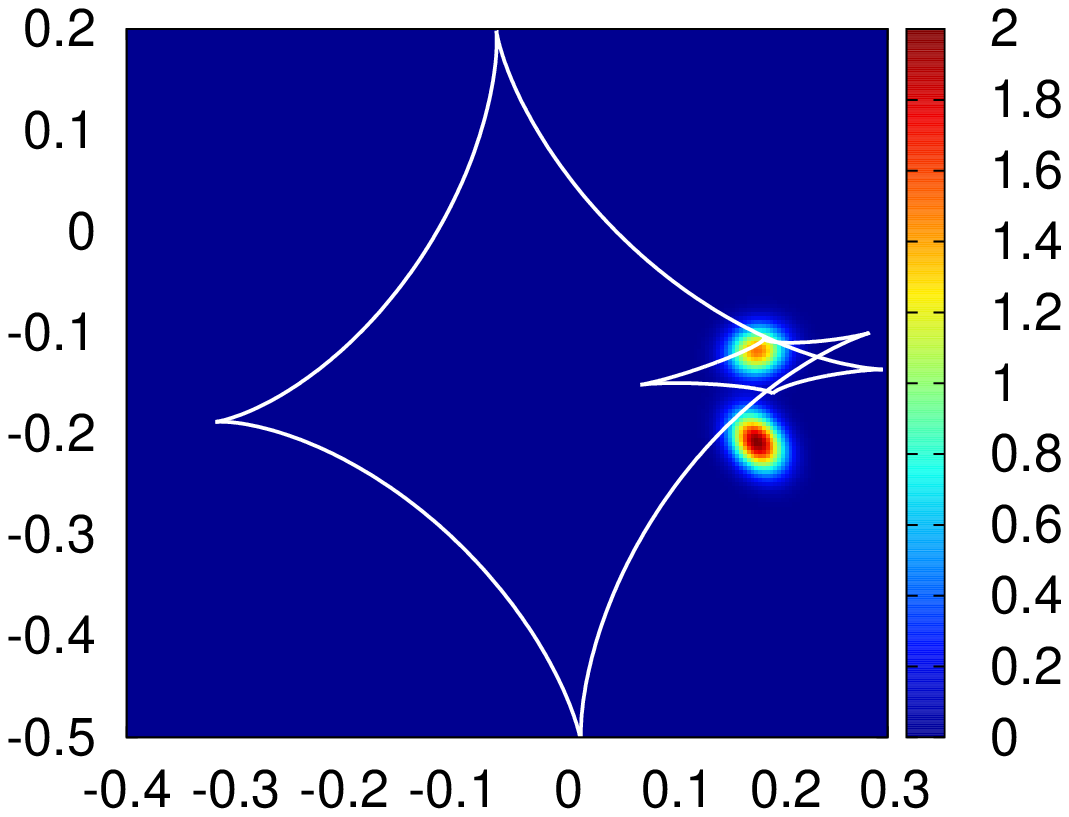}
		\label{fig:srcplane}
	}
	\subfigure[image plane]
	{
		\includegraphics[height=0.37\hsize,width=0.48\hsize]{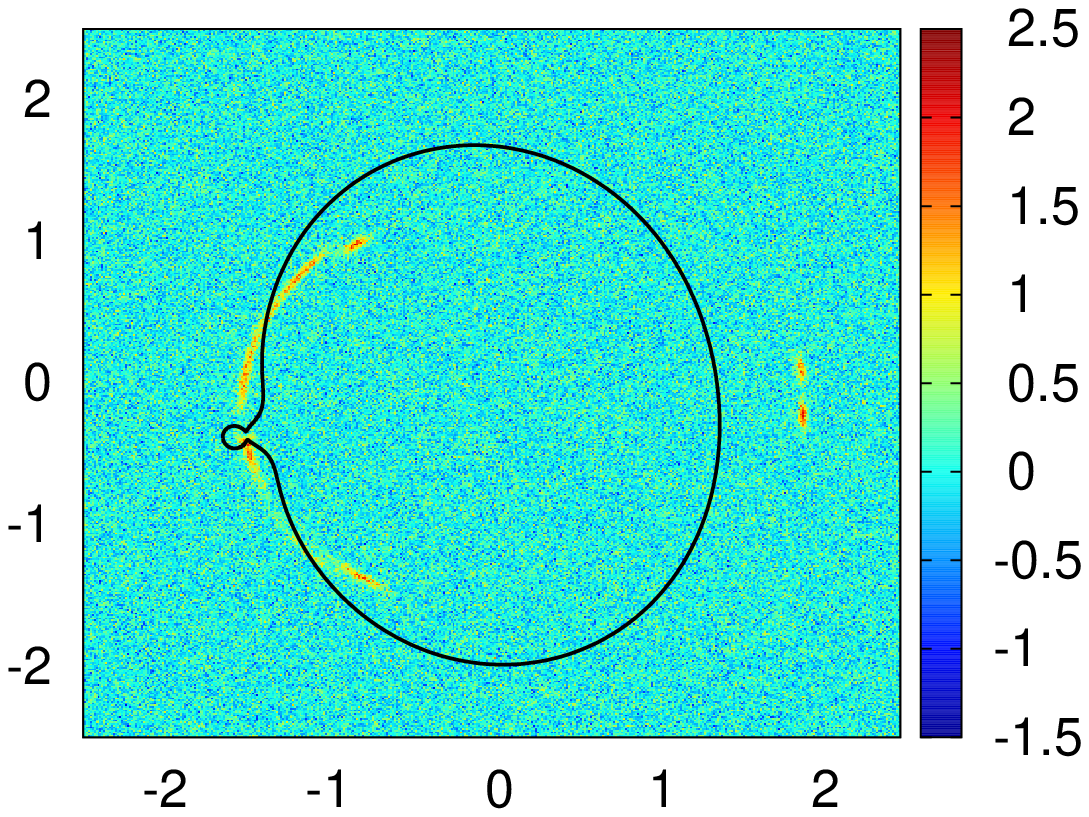}
		\label{fig:imgplane}
	}
	\caption{Simulated gravitational lens image where the source galaxy, host 
galaxy and external shear are similar to SDP.81 (axes are in arcseconds), and 
we have added a $\sim 10^{9}M_\odot$ subhalo centered at (-1.5,-0.37).  Here, 
we plot the first gravitational lens simulation listed in Table~\ref{tab:sims}, 
where the subhalo's inner slope is isothermal ($\gamma=2$) and the subhalo's 
distance to the host galaxy is equal to the Einstein radius of the lens 
($r_0=b$).  Critical curves and caustics are also plotted.}
	\vspace{10pt}
\label{fig:imgsrcplanes}
\end{figure*}

To investigate the effect of modeling a subhalo with an incorrect density 
profile, we simulate a gravitational lens similar to SDP.81 
\citep{alma2015,dye2015,wong2015,tamura2015,hatsukade2015}, wherein a detection 
of a $\sim 10^9M_\odot$ subhalo was recently reported (\citealt{hezaveh2016}; 
see also \citealt{inoue2016}).  The smooth lens component is modeled with an 
isothermal ellipsoid plus external shear, whose parameters are chosen to match 
those inferred for SDP.81 by \cite{rybak2015}.  The surface brightness profiles 
of the two primary structures in the reconstructed source galaxy are modeled 
with Gaussian profiles, and the simulated image is generated by ray-tracing 
plus the addition of Gaussian noise (see Figure \ref{fig:imgsrcplanes}). To be 
conservative, we choose a noise level such that the signal-to-noise ratio is 
approximately equal to 2 over the extent of the images.   In addition to adding 
Gaussian noise, we convolve the image with a Gaussian point spread function 
with dispersion 0.01 arcseconds (and thus a FWHM of $\approx$ 23 
milliarcseconds, comparable to ALMA's long-baseline resolution) along both the 
$x$ and $y$ directions. Our aim is not to perfectly simulate SDP.81, but rather 
to demonstrate the systematics of subhalo modelling in the context of a 
realistic gravitational lens. In particular, the correlated pixel noise arising 
from interferometry in actual ALMA data is not simulated here, since this is 
not the systematic we are interested in and would unnecessarily complicate the 
analysis.

\subsection{Subhalo density profile}

For the subhalo, we require a projected density profile with a variable 
log-slope and a smooth truncation at a specified tidal radius. To accomplish 
this, we choose the ``cuspy halo model'' of \cite{munoz2001}, whose density 
profile is

\begin{equation}
\rho = \frac{\rho_0 r_t^n}{r^\gamma \left(r^2+r_t^2\right)^{(n-\gamma)/2}}.
\label{eq:cuspyhalo}
\end{equation}

This model has an inner slope $\gamma$ and outer slope $n$. The Pseudo-Jaffe 
profile corresponds to the special case $\gamma=2$, $n=4$, so that the inner 
slope corresponds to that of an isothermal sphere, while the outer slope 
achieves a smooth truncation around the tidal radius $r_t$. For general 
$\gamma,n$ values, the projected (spherically symmetric) density profile 
$\kappa(r)$ and deflection $\alpha(r)$ have analytic solutions in terms of the 
Gaussian hypergeometric function as given in \cite{munoz2001}, which we use 
here.  As with the Pseudo-Jaffe profile, we will fix $n=4$ for the outer slope, 
but will allow the inner slope to vary.

We will also consider the effect of varying the subhalo distance from the host 
galaxy center (while keeping the projected position of the subhalo fixed). The 
resulting tidal radius of the subhalo can be estimated by considering the 
subhalo before truncation, whose density profile can be written as $\rho = 
\rho_0 (r_t/r)^\gamma$. Using the usual formula for the Jacobi radius, we find

\begin{equation}
r_t = \frac{r_0^2 \kappa_{s,0}}{(3-\gamma)b}
\label{rt_eq_corecusp}
\end{equation}
where $r_0$ is the distance from the subhalo to the center of the host galaxy, 
$b$ is the Einstein radius of the host, and $\kappa_{s,0} = 
2\pi\rho_0r_t/\Sigma_{cr}$ (our definition differs from that of 
\citealt{munoz2001} by a factor of $2\pi$).

For model fitting, we follow in the footsteps of previous work and model the 
subhalo explicitly with a Pseudo-Jaffe profile, which corresponds to 
$\gamma=2$, $n=4$ and whose projected density profile has a simple analytic 
form,

\begin{equation}
\kappa_s(r) = \frac{b_s}{2}\left[r^{-1} - \left(r^2+r_t^2\right)^{-\frac{1}{2}}\right].
\label{pjaffe_kappa}
\end{equation}

The parameter $b_s$ is related to $\kappa_{s,0}$ by $\kappa_{s,0} = b_s/r_t$; 
plugging this into Eq.~\ref{rt_eq_corecusp} gives $r_t = r_0\sqrt{b_s/b}$.  
Rather than vary $r_t$ explicitly, we will set $r_0 = b$, which then fixes $r_t 
= \sqrt{b b_s}$, recovering the usual formula for the tidal radius under this 
assumption.  Thus, our model subhalo parameters to be varied are $b_s$ along 
with the center coordinates $x_{c,s}$ and $y_{c,s}$.

\subsection{Simulated gravitational lens scenarios}

To investigate the effects of modeling a subhalo with an incorrect density 
profile, we consider four different gravitational lens simulations, which are 
listed and described in Table \ref{tab:sims}.  In each scenario, the ``actual'' 
subhalo density profile is given by Eq.~\ref{eq:cuspyhalo} and has the same 
projected position in the lens plane. For each simulated lens, we will follow 
the footsteps of previous work and fit the subhalo with a Pseudo-Jaffe model 
(Eq.~\ref{pjaffe_kappa}), whose tidal radius is fixed by the condition that its 
distance to the host galaxy center $r_0$ is equal to the Einstein radius $b 
\approx$ 1.6'' $\approx$ 7.4 kpc.  Compared to the model subhalo, the 
``actual'' subhalo density profile is identical in Simulation 1, has a 
shallower inner slope $\gamma$ in Simulations 2-3, and has a much larger 
distance from the host galaxy $r_0$=100 kpc in Simulation 4.  Note that a 
larger $r_0$ also implies a significantly larger the tidal radius $r_t$ 
according to Eq.~\ref{rt_eq_corecusp}.

\begin{figure}[t]
	\centering
	\subfigure[Sim~1 ($\gamma$=2, $r_t=0.22''$)]
	{
		\includegraphics[height=0.36\hsize,width=0.47\hsize]{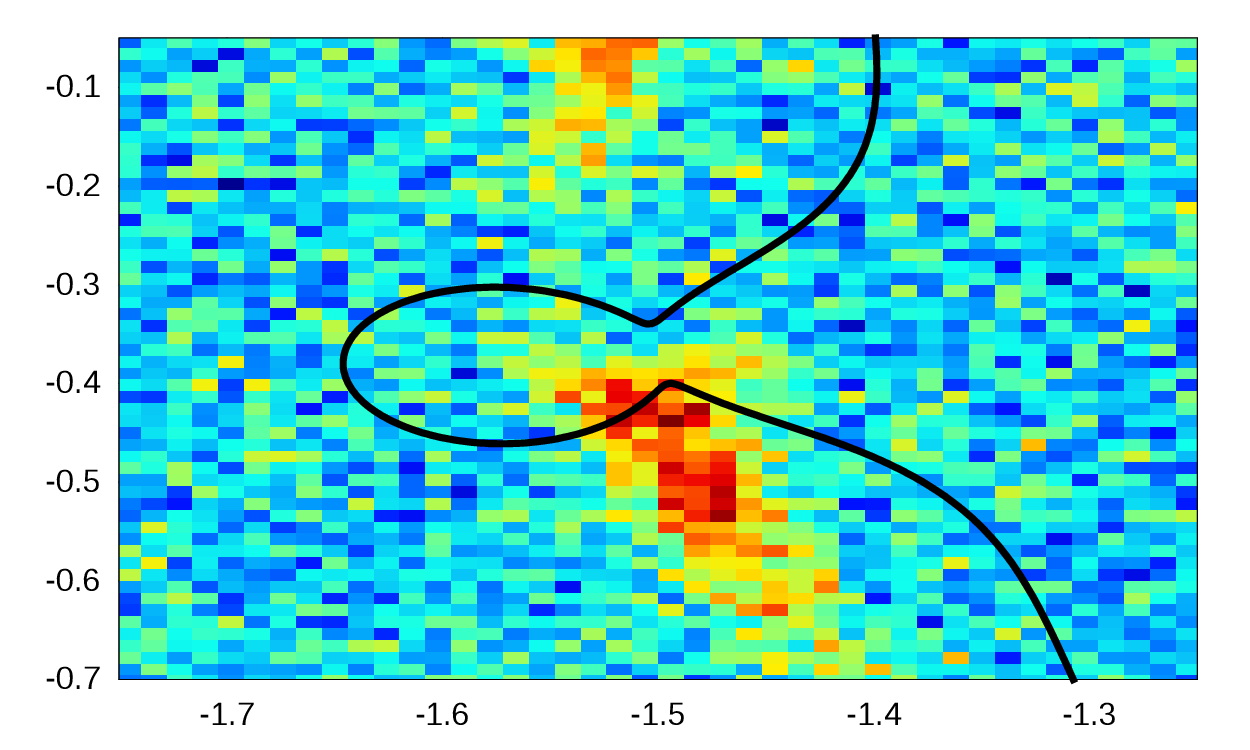}
		\label{pjsub}
	}
	\subfigure[Sim~2 ($\gamma$=1.5, $r_t=0.24''$)]
	{
		\includegraphics[height=0.36\hsize,width=0.47\hsize]{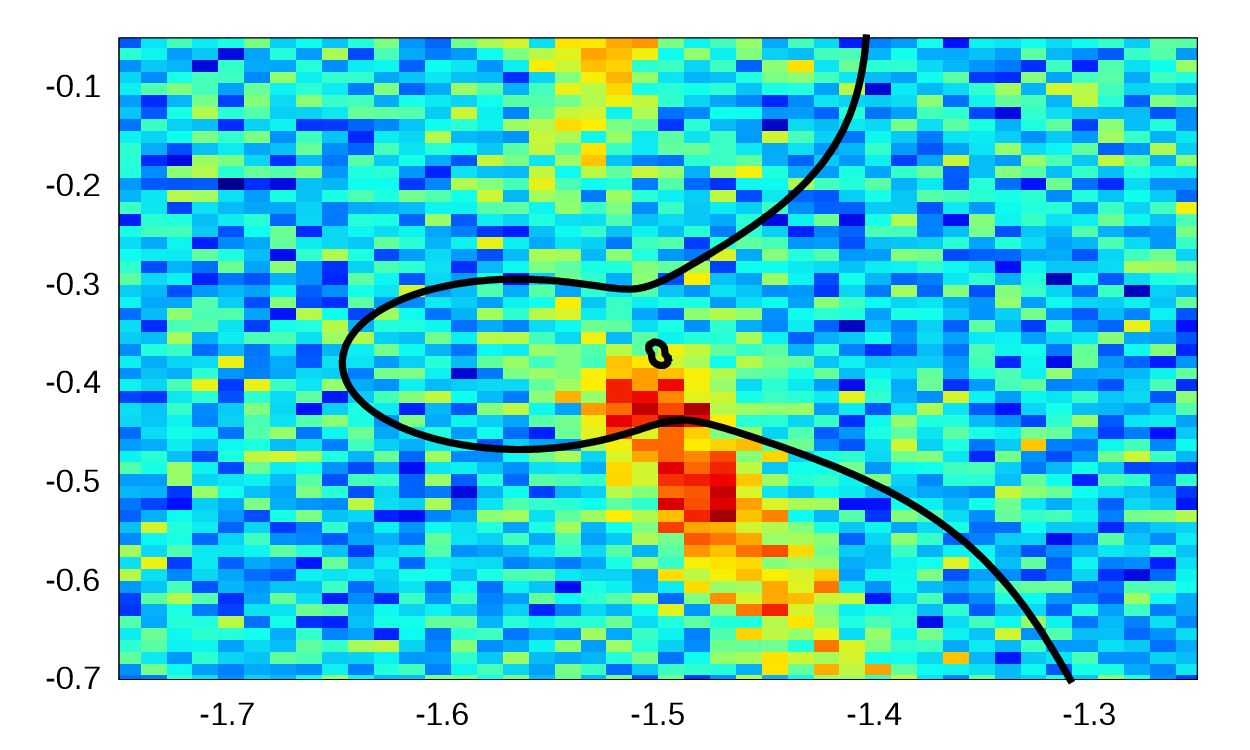}
		\label{cuspysub}
	}
	\subfigure[Sim~3 ($\gamma$=1, $r_t=0.27''$)]
	{
		\includegraphics[height=0.36\hsize,width=0.47\hsize]{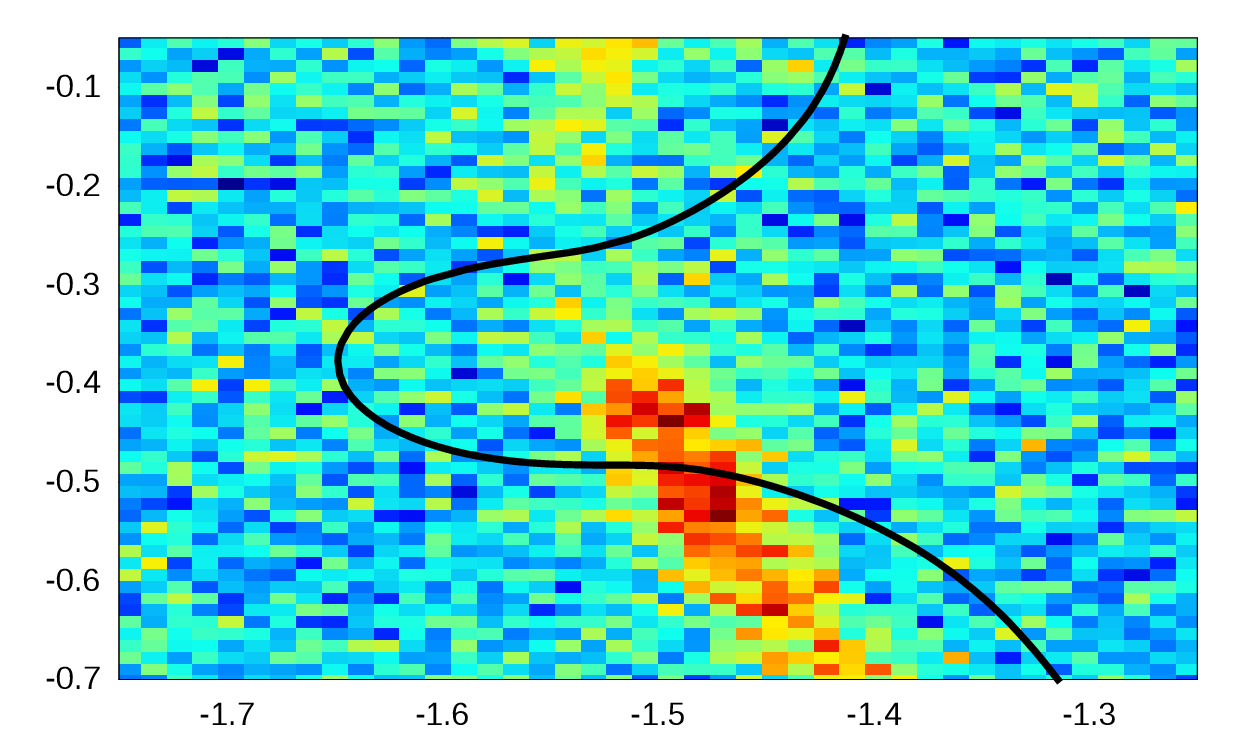}
		\label{cuspysub2}
	}
	\subfigure[Sim~4 ($\gamma$=2, $r_t=2.6''$)]
	{
		\includegraphics[height=0.36\hsize,width=0.47\hsize]{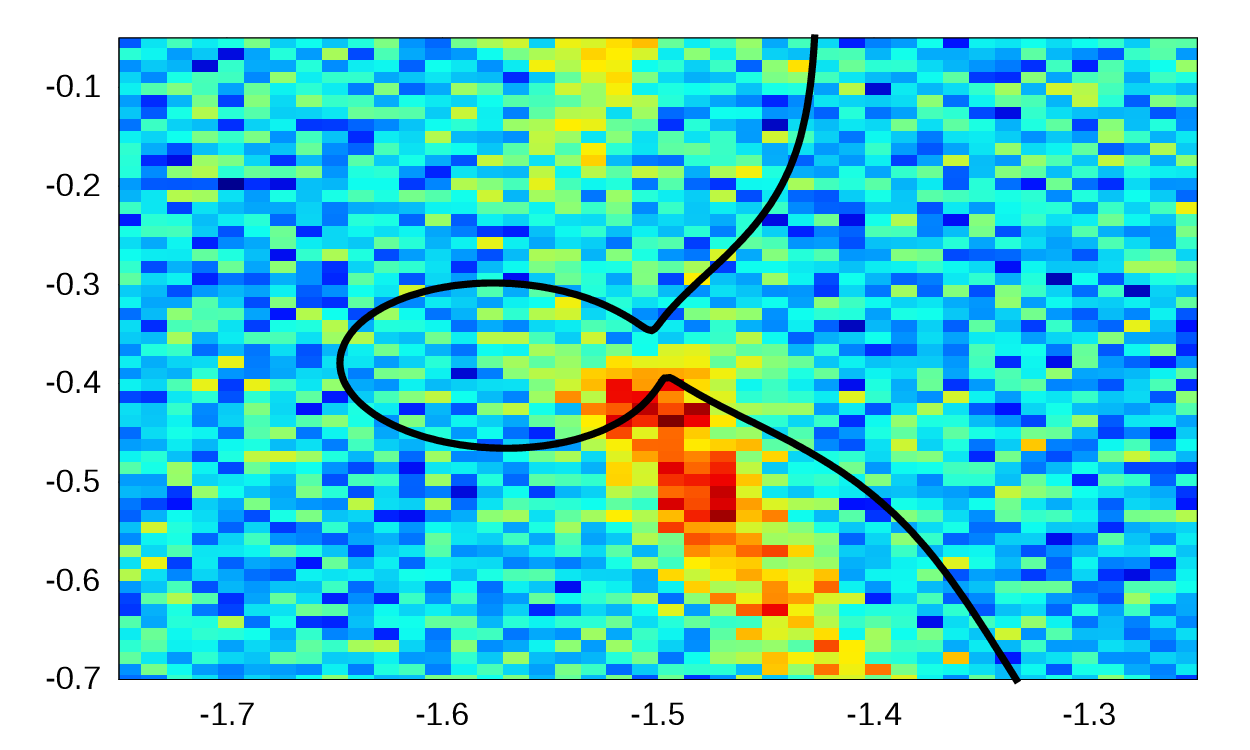}
		\label{pjsub100}
	}
	\caption{Zoomed-in region around the subhalo in each of the simulated 
gravitational lenses (which are described in Table~\ref{tab:sims}), with 
critical curves shown. Note that the subhalo perturbation radius is 
approximately the same in each simulation, i.e., the ``size'' of the critical 
curve perturbation is the same (by design). However, the shape of the critical 
curve perturbation differs dramatically: the critical curve is ``pinched'' more 
with a steeper subhalo profile, or with a larger tidal radius.}
\vspace{10pt}
\label{fig:zoom_imgs}
\end{figure}

The simulated image and source planes for Simulation 1 are shown in Figure 
\ref{fig:imgsrcplanes}. In the first simulation (which is our ``control'' 
scenario since the actual subhalo profile matches the model), we choose the 
normalization of the density profile $\kappa_{s,0}$ so that the total subhalo 
mass is approximately $10^9 M_\odot$. In the three remaining simulations, the 
normalization is chosen so that the mass within the subhalo perturbation radius 
$\delta\theta_{c}$ is roughly the same in each case, thus ensuring that the 
scale of the perturbation is not radically different in each case. Figure 
\ref{fig:zoom_imgs} plots the surface brightness and critical curves in each 
simulation. We see that indeed, the scale of the critical curve perturbation is 
approximately the same, although the shape of the perturbation differs: the 
steeper the subhalo density profile, the more the critical curve is 
``pinched''; likewise, the critical curve is pinched slightly more for a larger 
tidal radius.

For each scenario, we model the smooth lens component as a power-law ellipsoid 
\citep{tessore2015,barkana1998}, whose density profile can be expressed as

\begin{equation}
\kappa_0(x,y) = \frac{2-\alpha}{2}b^\alpha\left(qx^2 + 
y^2/q\right)^{-\frac{\alpha}{2}}
\label{power_law_ellipsoid_eq}
\end{equation}
where $q$ is the axis ratio. In this form, $b$ can be thought of as the average 
Einstein radius, since the semimajor axis of the critical curve is equal to 
$b/\sqrt{q}$ while the semiminor axis is equal to $b\sqrt{q}$.
Eq.~\ref{power_law_ellipsoid_eq} is then rotated by an angle $\theta$, where we 
define the $\theta$ to be the (counterclockwise) angle between the galaxy's 
semimajor axis and the $y$-axis of the observer's coordinate system.
Finally, in addition to the host galaxy, we add an external shear term which we 
parameterize by the shear components $\Gamma_1$, $\Gamma_2$.

\begin{table*}
\centering
\begin{tabular}{|l|c|c|c|c|c|c|c|c|c|c|c|c|}
\hline
& $b$ & $\alpha$ & $q$ & $\theta$ & $x_c$ & $y_c$ & $\Gamma_1$ & $\Gamma_2$ & $b_s$ & $x_{c,s}$ & $y_{c,s}$ & -$\log\mathcal{E}$\\
\hline
Actual values & 1.606 & 1 & 0.82 & 8.3$\degree$ & -0.019 & -0.154 & 0.0358 & 0.0038 & --- & -1.5 & -0.37 & --- \\
\hline
Simulation 1 & 1.606 & 1.018 & 0.814 & 8.623$\degree$ & -0.019 & -0.156 & 0.0368 & 0.0026 & 0.0321 & -1.495 & -0.372 & 13978 \\
Simulation 2 & 1.603 & 1.390 & 0.690 & 8.696$\degree$ & -0.045 & -0.160 & 0.0690 & 0.0085 & 0.0436 & -1.492 & -0.369 & 14004 \\
Simulation 3 & 1.606 & 1.394 & 0.680 & 9.829$\degree$ & -0.045 & -0.163 & 0.0663 & 0.0047 & 0.0504 & -1.473 & -0.367 & 14082 \\
Simulation 4 & 1.607 & 1.380 & 0.720 & 7.652$\degree$ & -0.045 & -0.157 & 0.0686 & 0.0110 & 0.0456 & -1.503 & -0.375 & 13989 \\
\hline
\end{tabular}
\caption{Best-fit model parameters for each simulated lens scenario (see 
Table~\ref{tab:sims} for a descriptive comparison of each simulation). The top 
row gives the actual parameters which are the same for all four simulations, 
except for $b_s$ since the subhalo model differs for each simulation. The 
remaining rows give the best-fit model parameters in each case. The final 
column gives the logarithm of the Bayesian evidence $\mathcal{E}$ for a 
goodness-of-fit comparison (\citealt{suyu2006}). }
\label{tab:bestfit}
\end{table*}

\subsection{Lens modeling algorithm}

To model our simulated lenses, we use the method of pixel-based source 
reconstruction, where the source is pixellated and the pixel values are 
inferred by linear inversion. Since this method is well-described in many 
papers (see e.g.~\citealt{tagore2014}, \citealt{warren2003}, 
\citealt{suyu2006}), we do not describe it in great detail in here, but rather 
point out distinct features of our algorithm.

For a given set of lens parameters, we use linear 3-point interpolation for the 
ray-tracing to construct the lensing matrix $L$, which is well-described in 
\cite{tagore2014}. We use curvature regularization for smoothing of the source, 
and perform a matrix inversion to obtain the inferred source surface brightness 
pixel map. The reconstructed source is pixellated using an adaptive Cartesian 
grid \citep{dye2005}, where the first-level pixels split into four smaller ones 
if the magnification of a source pixel is higher than $4\mu_0$, where $\mu_0$ 
is called the splitting threshold.  This process is then repeated recursively 
for the higher-level pixels. For simplicity, the number of first-level source 
pixels is fixed to $0.3$ times the number of image pixels being fit. (We note 
however that more sophisticated adaptive source grids are possible; see e.g.  
\citealt{vegetti2009} and \citealt{nightingale2015}.)

Apart from these points, we enumerate here the features that are unique in our 
implementation:

1)  In principle, the splitting threshold can be varied and the optimal value 
inferred from maximizing the Bayesian evidence. In practice we find that $\mu_0 
\approx 5-6$ is typically preferred, so to save computation time we fix the 
splitting threshold to $\mu_0=5$.

2) Although our adaptive source grid is Cartesian, its borders are not fixed, 
but instead are redrawn each time the lens parameters are varied so that it is 
just large enough to include the inferred source signal. In this way, we 
exclude as many pixels as possible that only contain noise which could degrade 
the fit. In practice, we accomplish this by drawing the box so that it just 
barely contains all pixels for whom the surface brightness is greater than 1.3 
times the pixel noise. To be conservative, we then extend the length and width 
of the box by 15\% to make sure all the signal is being included. A side 
benefit of this is that computation time is saved, since typically fewer pixels 
are being evaluated compared to a fixed grid.

3) As is common practice, we mask regions of the image which only contain noisy 
pixels, so that the majority of the pixels being fit to contain the actual 
signal.  However, there is a risk that certain solutions produce images in the 
masked region (i.e.  outside the region being fit) that should not exist, and 
this may be missed.  To avoid this, after inverting to find the source surface 
brightness distribution, we then reconstruct the lensing matrix using all of 
the pixels in the image, including the masked regions, and find the resulting 
surface brightness in each pixel. We then impose a penalty function prior, so 
that if the surface brightness in the masked region rises above 30\% of the 
brightest pixel in the image, a steep penalty is imposed and this solution is 
discouraged.

4) In subhalo modeling studies that have used pixellated source reconstruction 
to date, typically an nonparametric approach \citep{hezaveh2016,vegetti2009} is 
first used to attempt to locate where the lensing potential is being perturbed 
by a subhalo, before proceeding to the full parameterized subhalo model. One 
very important reason for doing so, rather than a parametric fit alone, is that 
if the mass distribution of the smooth lens component is more complicated than 
the chosen parametric model, spurious detections can result in an attempt to 
improve the fit \citep{vegetti2014b}.  In this paper, since we are working with 
simulated data, we skip this step entirely; from a practical standpoint, this 
is possible thanks to an adaptive Markov Chain Monte Carlo (MCMC) algorithm 
called T-Walk \citep{christen2010}, which is uniquely suited to efficient 
exploration of multimodal distributions.  The T-Walk algorithm uses a 
Metropolis-Hastings step, but at each step, certain chains perform a ``jump'' 
over the current point in another, randomly chosen chain in an attempt to find 
new modes in the posterior distribution. In addition, linear coordinate 
transformations are performed in order to better handle parameter degeneracies.  
While a substantial number of chains may be required to confidently sample the 
parameter space, we find the T-Walk algorithm nevertheless converges much 
faster than a nested sampling algorithm even if there are many separate modes 
in the parameter space. One important benefit is that if there are multiple 
subhalo perturbations in the lensed image, the MCMC can find both modes even if 
only one subhalo is included in the model. One disadvantage of using T-Walk, 
however, is that it is not necessarily clear ahead of time how many chains will 
be needed to sample all the modes in the parameter space---for this reason 
along with the reasons stated above, a nonparametric analysis to identify lens 
perturbations in actual data is still recommended before proceeding to the 
MCMC.

\begin{figure*}[p!]
	\centering
	\subfigure[Sim 1, data]
	{
		\includegraphics[height=0.34\hsize,width=0.48\hsize]{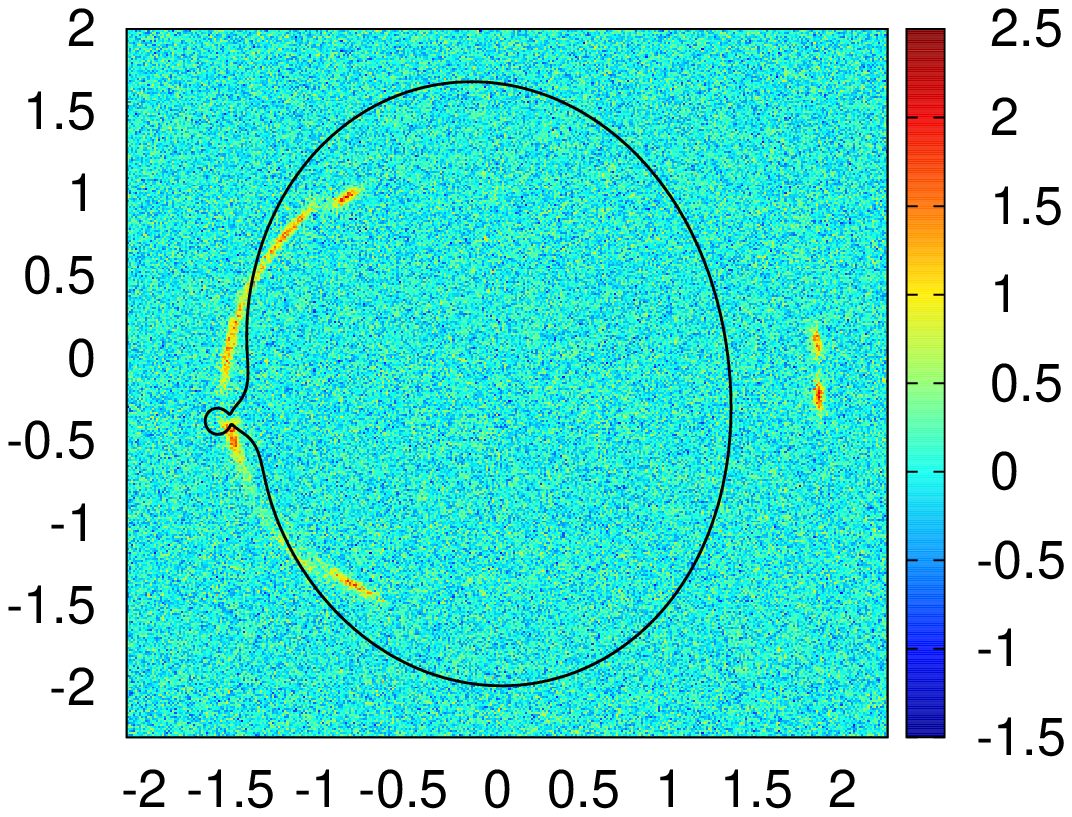}
		\label{fig:sim1data}
	}
	\subfigure[Sim 1, best-fit solution]
	{
		\includegraphics[height=0.34\hsize,width=0.48\hsize]{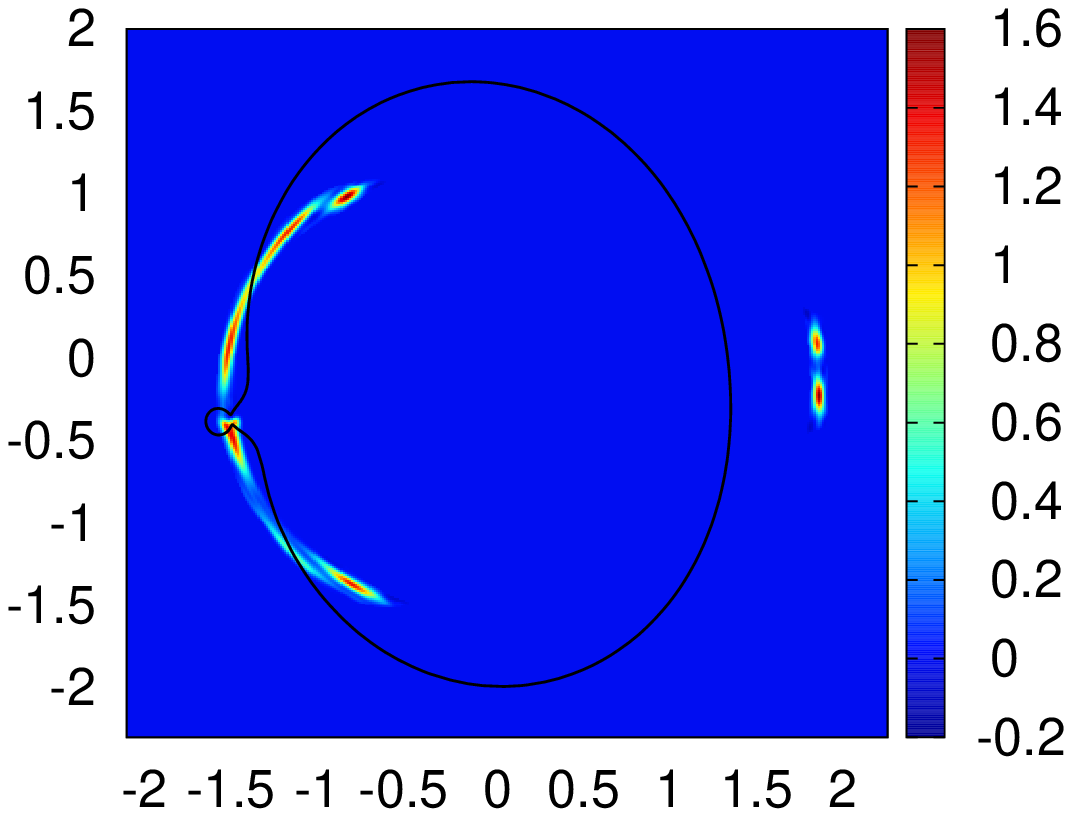}
		\label{fig:sim1fit}
	}
	\subfigure[Sim 3, data]
	{
		\includegraphics[height=0.34\hsize,width=0.48\hsize]{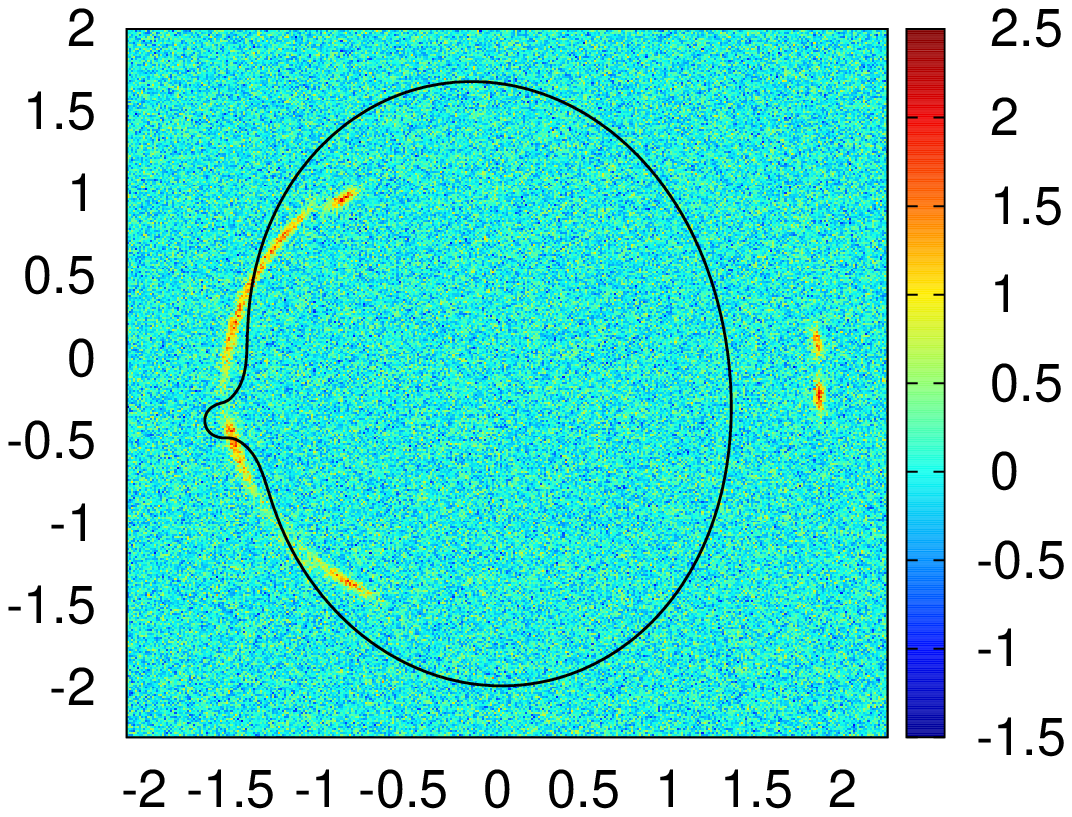}
		\label{fig:sim3data}
	}
	\subfigure[Sim 3, best-fit solution]
	{
		\includegraphics[height=0.34\hsize,width=0.48\hsize]{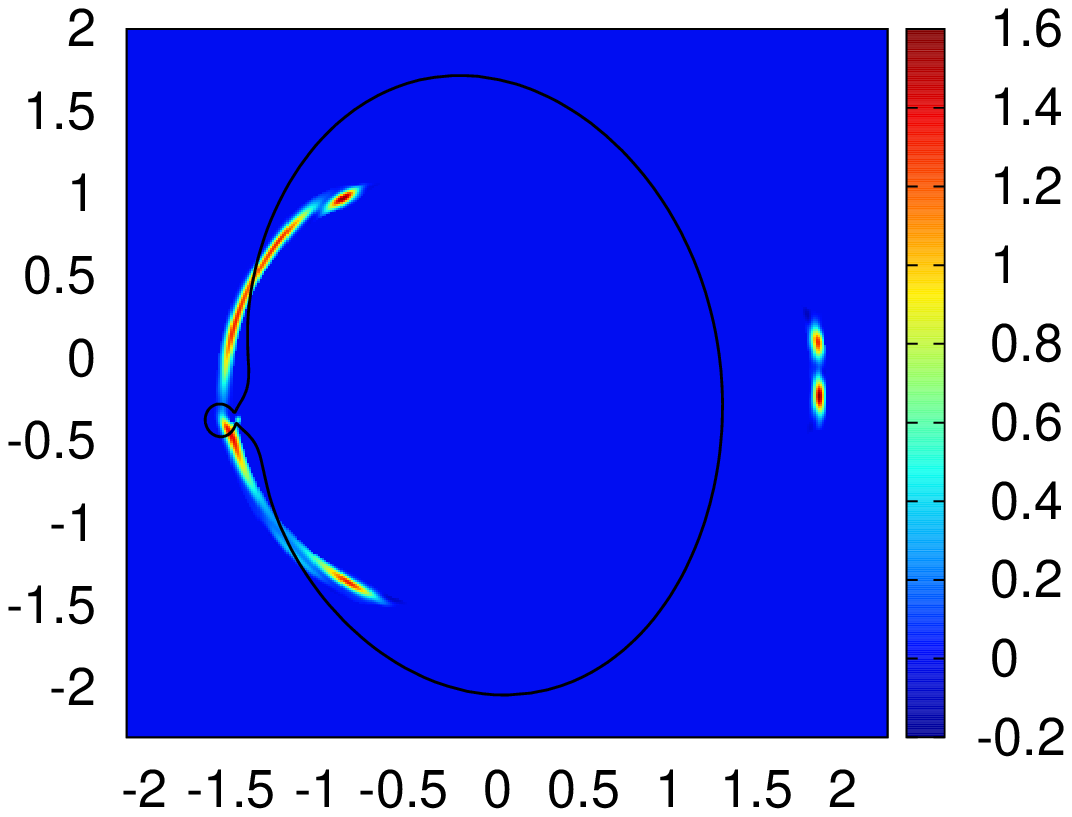}
		\label{fig:sim3fit}
	}
	\caption{Pixel surface brightness data for Simulations 1 and 3, along with 
best-fit reconstructions and critical curves. (See Table \ref{tab:sims} for a 
descriptive comparison of each simulation.) Note that in the reconstructed 
image plane for Simulation 3, a small additional image is present near the 
position of the subhalo which does not appear in the data; this is a 
consequence of modeling the subhalo with a steeper profile compared to the 
actual subhalo. Despite this, although the shape of the critical curve 
perturbation is different compared to the data, the scale of this perturbation 
$\delta\theta_c$ is remarkably well-reproduced.}
	\vspace{10pt}
\label{fig:simfits}
\end{figure*}

\begin{figure*}[p!]
	\centering
	\subfigure[Sim 1, best-fit reconstructed source]
	{
		\includegraphics[height=0.30\hsize,width=0.45\hsize]{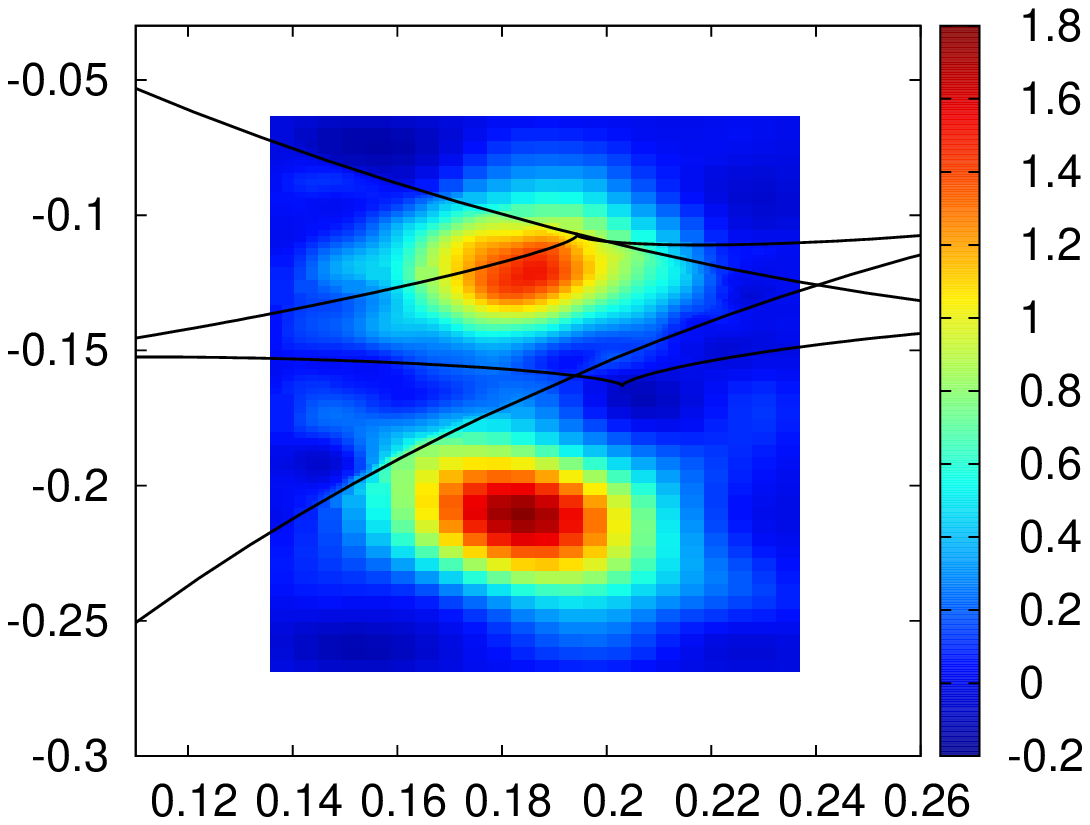}
		\label{fig:sim1src}
	}
	\subfigure[Sim 3, best-fit reconstructed source]
	{
		\includegraphics[height=0.30\hsize,width=0.45\hsize]{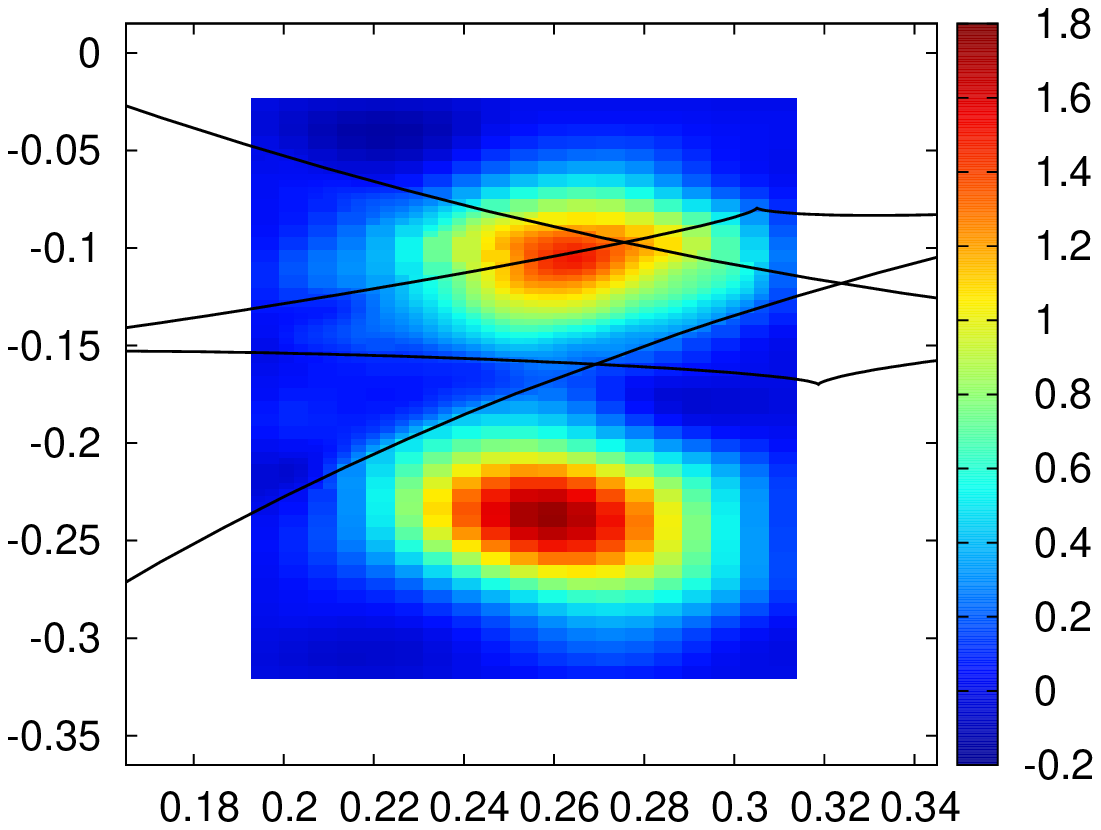}
		\label{fig:sim3src}
	}

	\caption{Reconstructed source plane for Simulations 1 and 3. The actual 
source surface brightness distribution is the same for all the simulations and 
shown in Figure \ref{fig:srcplane}. Note that source pixels are recursively 
split into 4 smaller subpixels in regions of high magnification.  Despite the 
differences in each simulated lens image and reconstruction, the source 
structure is well-reproduced in each case, although the position and scale of 
the source pixels vary somewhat between the different simulations.}
	\vspace{10pt}
\label{fig:simsrcs}
\end{figure*}

\begin{figure*}
	\centering
	\includegraphics[height=1.0\hsize,width=1.0\hsize]{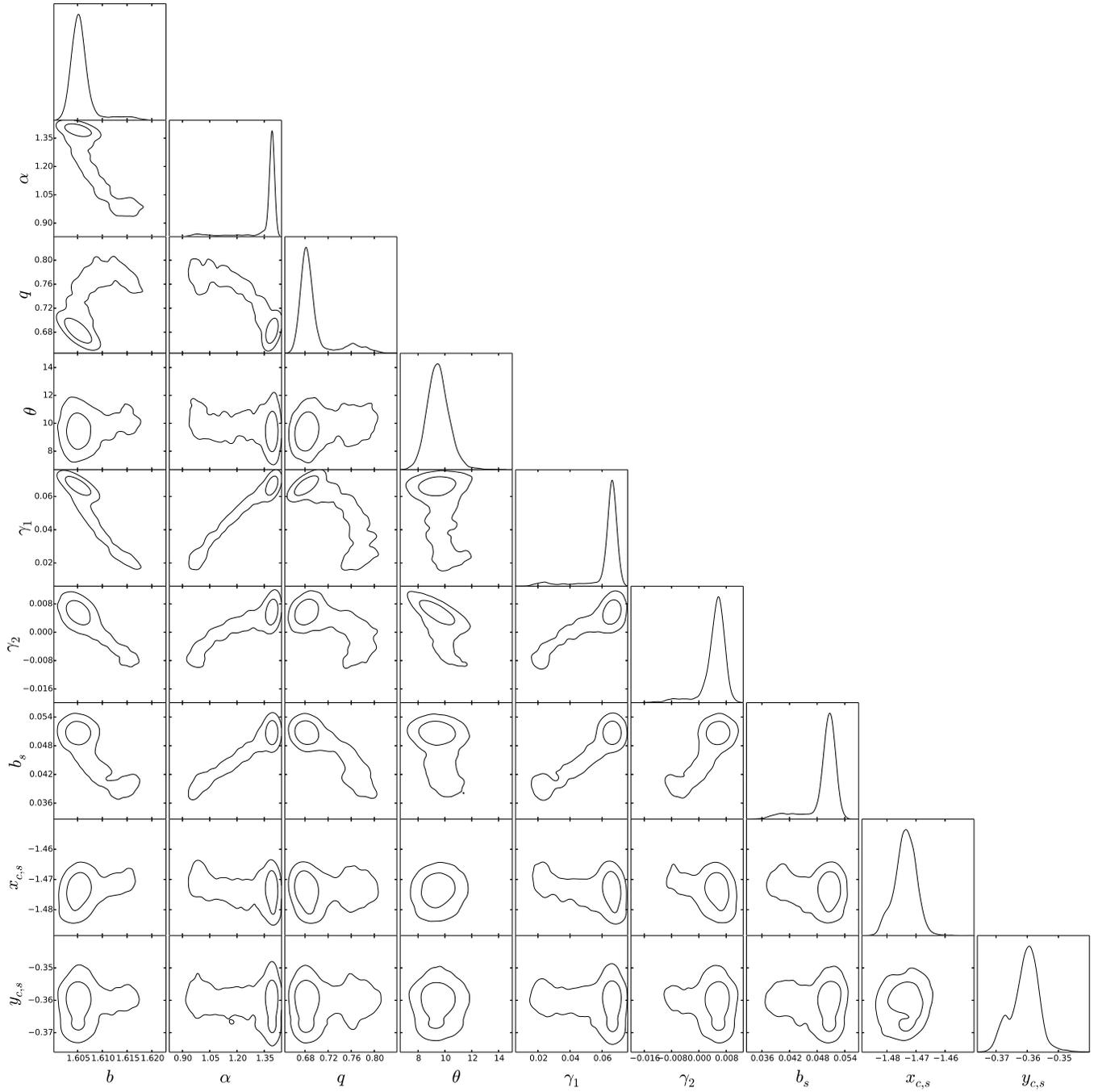}
\caption{Posterior probability distributions in the parameters for Simulation 3,
where the subhalo has an inner log-slope -1 (see Table~\ref{tab:sims} for description). To reduce 
clutter, we have not shown the posteriors for the position of the host galaxy 
($x_c$,$y_c$) or the regularization parameter, which are very well-constrained.  
Note that the log-slope of the host galaxy $\alpha$ is biased high, which
attempts to mitigate the overly ``sharp'' perturbation created by the steeper density profile of the
subhalo (inner log-slope -2) in the model.}
\label{fig:triangle_plot}
\end{figure*}

\begin{table*}
\centering
\begin{tabular}{|l|c|c|c|c|c|c|c|}
\hline
 & $m_{T}/10^9M_\odot$ & $m_{T,fit}/10^9M_\odot$ & $\Delta m_T/m_T$ & $\delta\theta_{c}$ (kpc) & $\tilde m_{sub}/10^9M_\odot$ & $\tilde m_{sub,fit}/10^9M_\odot$ & $\Delta \tilde m_{sub}/\tilde m_{sub}$ \\
\hline
Simulation 1 & 1.027 & 1.058 & 3.02\% & 0.679 & 0.473 & 0.476 & 0.60\% \\
Simulation 2 & 1.504 & 1.674 & 11.3\% & 0.716 & 0.494 & 0.518 & 4.87\% \\
Simulation 3 & 2.341 & 2.100 & 10.3\% & 0.809 & 0.592 & 0.662 & 11.7\% \\
Simulation 4 & 8.698 & 1.794 & 79.4\% & 0.744 & 0.519 & 0.556 & 7.12\% \\
\hline
\end{tabular}
\caption{Comparison of subhalo mass estimators for the actual vs. best-fit 
values in each simulated gravitational lens (see Table~\ref{tab:sims} for a 
descriptive comparison of each simulation). $m_{T}$ refers to the total subhalo 
mass, while the effective subhalo mass $\tilde m_{sub} \equiv 
m_{sub}(\delta\theta_c)/\alpha$ is the projected subhalo mass enclosed within 
the subhalo perturbation radius $\delta\theta_c$ (which we list in parsecs in 
the middle column).  Note that for Simulation 4, where the subhalo has a 
substantially larger tidal radius, the inferred total subhalo mass $m_T$ is off 
by $\sim 80$\% while the effective subhalo lensing mass $\tilde m_{sub}$ is 
recovered very well by comparison.}
\label{tab:mass_comparison}
\end{table*}

\begin{figure*}[t]
	\centering
	\subfigure[Simulation 1 ($\delta\theta_c \approx 0.149$'')]
	{
		\includegraphics[height=0.35\hsize,width=0.48\hsize]{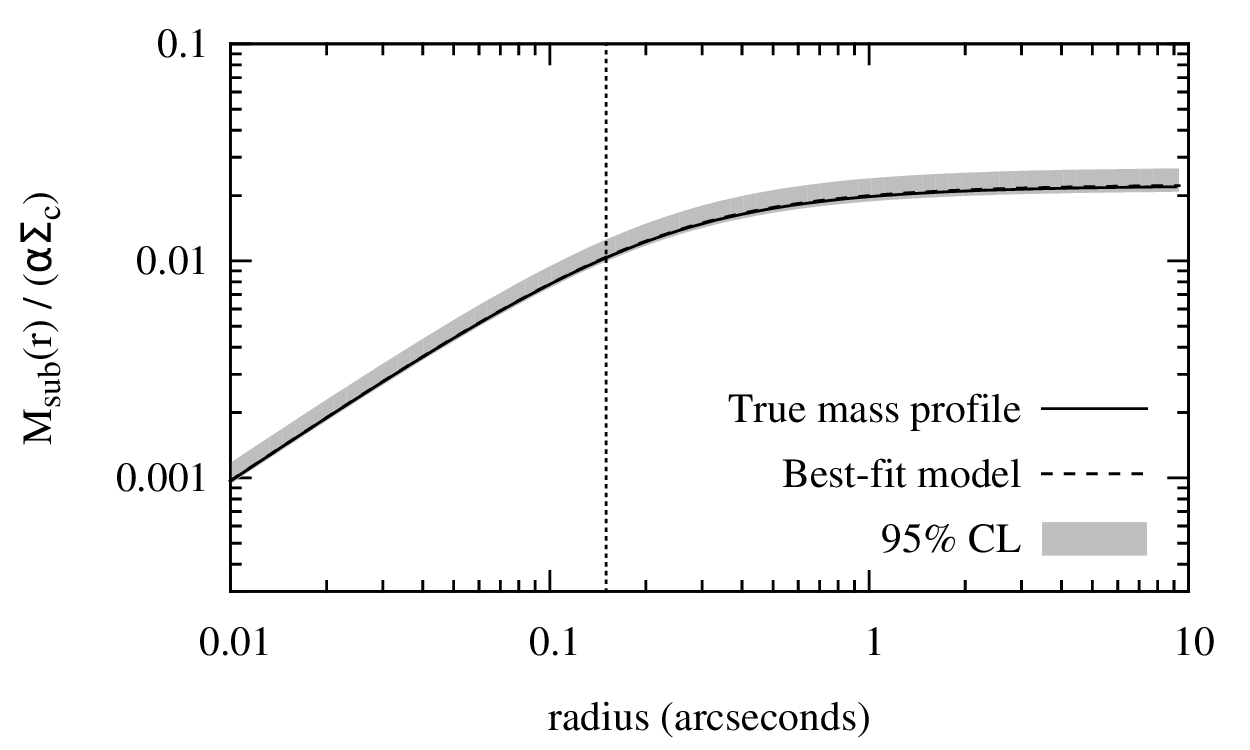}
		\label{mprofile_pjsub}
	}
	\subfigure[Simulation 2 ($\delta\theta_c \approx 0.155$'')]
	{
		\includegraphics[height=0.35\hsize,width=0.48\hsize]{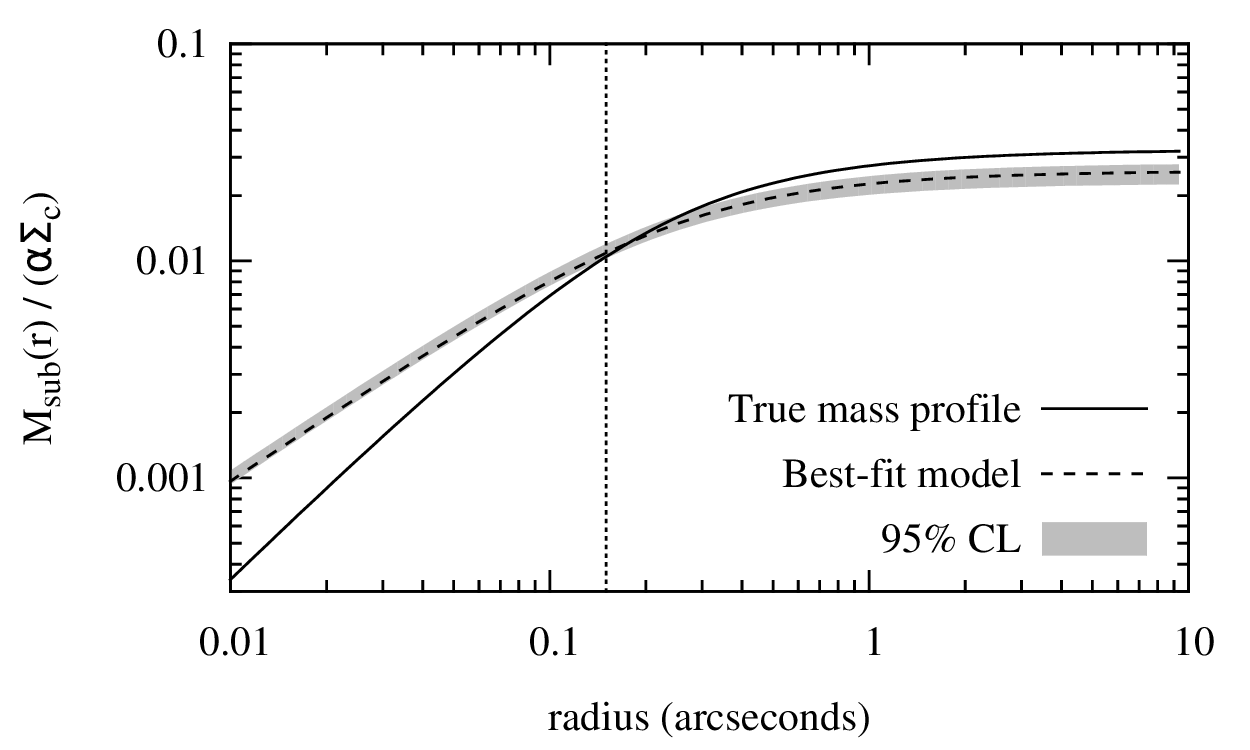}
		\label{mprofile_cuspysub}
	}
	\subfigure[Simulation 3 ($\delta\theta_c \approx 0.175$'')]
	{
		\includegraphics[height=0.35\hsize,width=0.48\hsize]{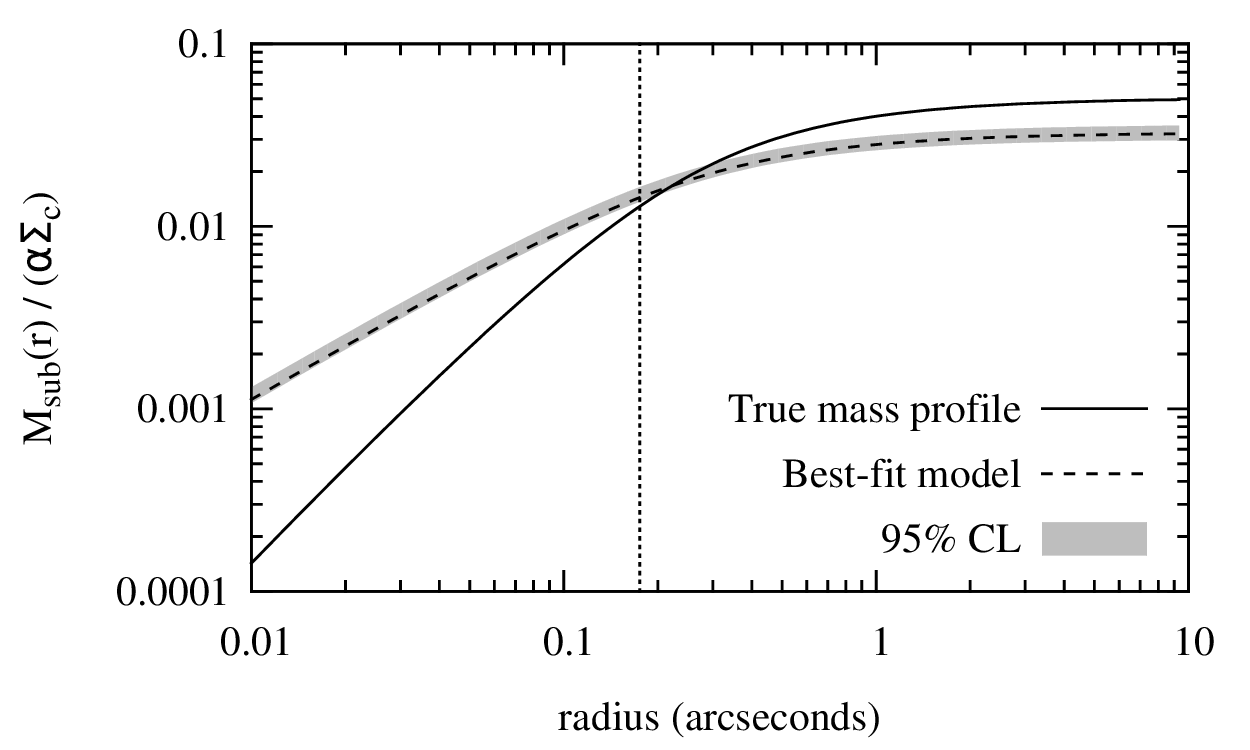}
		\label{mprofile_cuspysub2}
	}
	\subfigure[Simulation 4 ($\delta\theta_c \approx 0.161$'')]
	{
		\includegraphics[height=0.35\hsize,width=0.48\hsize]{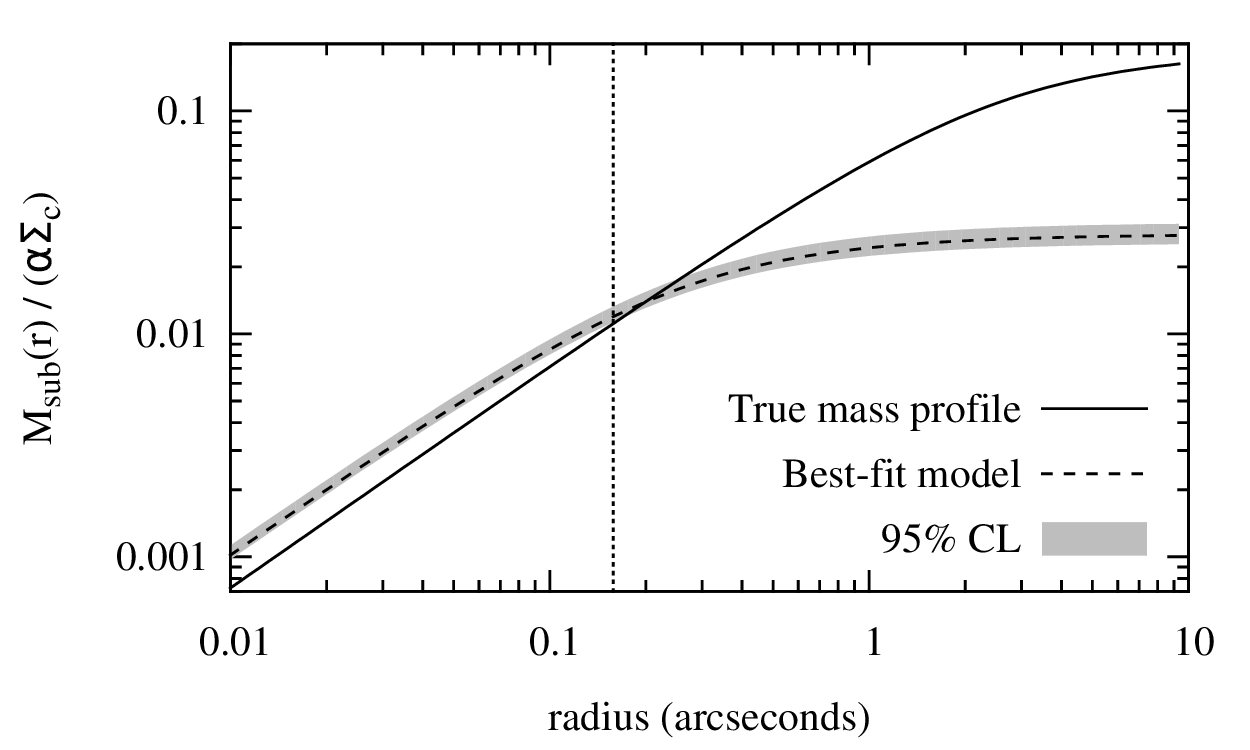}
		\label{mprofile_pjsub100}
	}
	\caption{Best-fit mass profile of the subhalo compared to the true profile 
in each simulated lens, scaled by $1/\alpha\Sigma_{cr}$ (where $\alpha$ is the 
log-slope of the host galaxy's density profile). For a descriptive comparison 
of each simulation, see Table~\ref{tab:sims}. The vertical dashed line denotes 
the subhalo perturbation radius $\delta\theta_{c}$, which is the radius where 
the critical curve is maximally perturbed by the subhalo in the best-fit model; 
the best-fit $\delta\theta_c$ is given below each plot for comparison.  In all 
cases except for Simulation 1, the total mass of the best-fit model is biased 
due to the mismatched density profile; nevertheless the mass enclosed within 
$\delta\theta_{c}$ is well-recovered, provided it is scaled by $1/\alpha$.}
	\vspace{10pt}
\label{fig:mprofiles}
\end{figure*}

\section{Results}\label{sec:results}

After running the T-Walk MCMC algorithm to explore the parameter space, 
posteriors are generated in each parameter; we plot the resulting joint 
posteriors for Simulation 3 in Figure \ref{fig:triangle_plot}.  Finding the 
correct parameter space for the subhalo required a substantial number of chains 
(up to 96 chains). In the process, the chains found several false modes in 
parameter space with the subhalo located incorrectly, that partially mimicked 
the effect of the subhalo.  Typically these false modes involved at least one 
extra source structure; for example, in one mode, one of the actual source 
structures was split into two, each of which were highly magnified.  These 
kinds of modes involved a greater number of subpixels and/or greater 
regularization, and hence were disfavored by the Bayesian evidence compared to 
the ``correct'' mode. This highlights a major advantage of using the Bayesian 
evidence (beyond just setting the optimal regularization, which is discussed 
extensively in \citealt{suyu2006}) rather than simply the likelihood.

The resulting best-fit parameters in each case are shown in Table 
\ref{tab:bestfit}. For reference, we plot the reconstructed image and 
source planes for Simulations 1 and 3 in Figures \ref{fig:simfits} and \ref{fig:simsrcs} 
respectively. As expected, the model parameters in Simulation 1 are 
well-recovered, since the subhalo density profile is the same as in the model.  
In the other simulations, the position of the subhalo is fairly well-recovered, 
but some of the host halo parameters are biased in an attempt to make up for 
the fact that the subhalo in the model does not perfectly reproduce the 
observed perturbation.  Most prominently, the model prefers the log-slope of 
the host galaxy's profile $\alpha$ to be greater than 1; in fact, the best-fit 
values of $\alpha$ were close to our upper prior limit of 1.4.  We believe this 
effect is a consequence of the shallower projected density profile in the 
actual subhalo in each simulation.  In these cases, as Figure 
\ref{fig:zoom_imgs} shows, the subhalo perturbation is less ``sharp'' and does 
not pinch the critical curve as dramatically compared to the Pseudo-Jaffe 
subhalo; as a result, the magnification is lower in the vicinity of the 
subhalo. One consequence of this is that a small additional image is evident 
just to the right of the subhalo in Simulation 1 (Figures \ref{fig:sim1data}, 
\ref{fig:sim1fit}), whereas for Simulations 2-3 which have shallower profiles, 
these additional images are not evident (see Figure \ref{fig:sim3data} for 
Simulation 3).  Nevertheless, in the reconstructed solution, a small additional 
image does appear (Figure \ref{fig:sim3fit}), which is inevitable because we 
are modeling with a Pseudo-Jaffe subhalo. It seems that having a steeper 
density profile for the host galaxy mitigates this effect by reducing the 
magnification and hence the size of the additional image produced by the 
Pseudo-Jaffe subhalo, and hence a steeper log-slope $\alpha$ is preferred.  
However, this does not mean that $\alpha=1$ was excluded in all cases; for 
example, in Simulation 3, the posterior extends all the way down to 
$\alpha\approx 1$ (see Fig.~\ref{fig:triangle_plot}), although the fit is not 
as good there.

\subsection{How biased is the inferred mass of the subhalo?}

To investigate the bias in the inferred subhalo mass in each scenario, in Table 
\ref{tab:mass_comparison} we show the ``true'' total subhalo mass and the 
best-fit subhalo mass for each scenario. As expected, the total subhalo mass is 
well-recovered in Simulation 1, since the subhalo density profile matches that 
of the model. In Simulations 2 and 3 where the slope of the density profile 
differs but the subhalo distance matches the model ($r_0=b$), the error is 
still fairly small, roughly 11\% in each case. This is a fortunate consequence 
of the fact that the scale of subhalo lensing perturbation is naturally similar 
to the tidal radius under this assumption, as discussed in Section 
\ref{sec:subhalo_perturbation}.  However, the situation is very different if 
the tidal radius differs significantly from the model, as is evident in the 
Simulation 4 results: here, the actual subhalo mass is nearly five times larger 
than the inferred mass, with an error of nearly 80\%. The reason is clear: in 
this case the subhalo's tidal radius is approximately 2.6 arcseconds, compared 
to 0.27 arcseconds in the best-fit model. Thus, the subhalo's density profile 
extends much further out compared to the model, and beyond the critical curve 
perturbation radius $\delta\theta_{c}$, this extra mass can be ``hidden'' by 
adjusting the smooth lens component. This high systematic error in the total 
subhalo mass has also been pointed out by \cite{vegetti2012} and 
\cite{vegetti2014} (indeed they estimate this error by considering different 
projected distances to the lens plane, under the assumption that the subhalo 
number density traces the dark matter profile of the host galaxy).

It has been suggested in \cite{vegetti2014} that the mass enclosed within the 
subhalo's Einstein radius is robust to changes in the density profile, so we 
also test this idea on our simulated lenses. With the exception of Simulation 1 
(where the subhalo's density profile and tidal radius are correctly modeled), 
we find that at the subhalo's Einstein radius $b_s$ in the best-fit model 
(which is in the range 0.04-0.05''; see Table~\ref{tab:bestfit}), the enclosed 
subhalo mass is significantly overestimated. In the most extreme case, 
Simulation 3, the best-fit subhalo mass enclosed within its Einstein radius is 
$\sim$ 3.5 times larger than the actual mass, an error of 250\%. It is also 
worth noting that the Einstein radius of the actual subhalo can also be 
dramatically different from the model if the slope of the density profile 
differs enough; in Simulation 3, the true Einstein radius of the subhalo is 
$7.3\times 10^{-4}$ arcseconds, much smaller than the Einstein radius of the 
best-fit model subhalo.  Thus, we conclude that the Einstein radius of the 
subhalo itself plays no special role and the inferred mass within this radius 
is not robust to changes in the density profile. 

\subsection{Robustness of the effective subhalo lensing mass}

Having established that the bias in the inferred subhalo mass can be 
significant, we now investigate whether the effective subhalo lensing mass 
$\tilde m_{sub} \equiv m_{sub}(\delta\theta_{c})/\alpha$ (defined in Section 
\ref{sec:subhalo_perturbation}) can indeed be considered a robust mass 
estimator.  In Figure \ref{fig:mprofiles} we plot the subhalo mass profile of 
the best-fit model in each case (dashed line) divided by the log-slope $\alpha$ 
of the host galaxy, compared to the actual subhalo mass profile (solid line; 
recall that $\alpha=1$ for the actual model in each case).  The shaded region 
denotes the 95\% confidence interval, which is obtained by calculating a 
derived posterior probability in the subhalo mass enclosed within each radius.  
Simulation recovers the mass profile very well, as expected; for the remaining 
cases, the mass enclosed within $\delta\theta_{c}$ (denoted by the dotted 
vertical line) divided by $\alpha$ is indeed quite close to the actual subhalo 
mass even though the mass elsewhere is different. In Table 
\ref{tab:mass_comparison} the values for $\tilde m_{sub}$ are tabulated along 
with the error. In Simulation 4, the error in this quantity is roughly 8\%, a 
vast improvement compared to the discrepancy in the total subhalo mass.  With 
the exception of the shallowest density profile $\gamma=1$ (Simulation 3), the 
error in $\tilde m_{sub}$ is significantly smaller than the error in the total 
subhalo mass.

Further evidence for the invariance of the effective subhalo mass $\tilde 
m_{sub}$ can be seen in the posteriors themselves. In the parameter region 
where the data are well-fit, if
$m_{sub}(\delta\theta_{c})/\alpha$ is invariant then it follows that 
$b_s/\alpha$ must also be approximately invariant. Indeed, this can be seen in 
the approximate formula Eq.~\ref{rmax_approx_eq2}: to highest order, 
$\delta\theta_c$ will remain the same if the ratio $b_s/\alpha$ is fixed.  Thus 
we expect a linear degeneracy between the parameters $b_s$ and $\alpha$, which 
is shown beautifully in the joint posterior in these parameters in Figure 
\ref{fig:triangle_plot}.  For the best-fit solution, we have $b_s \approx 
0.05$, $\alpha\approx 1.4$, and thus the ratio $b_s/\alpha \approx 0.036$. If 
we pick a different point, for example where $\alpha=1.2$, then to maintain 
this ratio we must therefore have $b_s \approx 0.036\times1.2 \approx 0.043$, 
and indeed this value falls within the allowed parameter region in the figure.  
The relationship is not satisfied quite as well for $\alpha=1$, where the 
posterior lies slightly above the expected value $b_s\approx 0.036$; however, 
in this region the subhalo position has drifted further away from the critical 
curve, which can be seen in the joint $\alpha$-$x_{c,s}$ posterior at 
$\alpha=1$.  Since the subhalo position has shifted, the effective subhalo 
lensing mass will change, and in order to produce the same critical curve 
perturbation $b_s$ needs to be slightly larger, as the posterior shows.  
Nevertheless, in the neighborhood of the best-fit region it is clear that the 
$\alpha$-$b_s$ degeneracy is fully consistent with the invariance of the 
effective subhalo lensing mass $\tilde m_{sub}$.

\subsection{Sources of bias in the effective subhalo lensing mass}

Although Figure \ref{fig:mprofiles} shows that the effective subhalo lensing 
mass $\tilde m_{sub}$ is fairly robust, it is also biased a little bit high in 
each case.  Why is this?  There are two reasons.  First, although 
$\delta\theta_{c}$ in the best-fit model is close to the actual value in each 
case, it is not \emph{exactly} the same---it differs, either because the 
critical curve perturbation itself is larger or the subhalo is slightly farther 
from the unperturbed critical curve compared to the actual subhalo.  This is 
most evident in Simulation 3, where the best-fit subhalo position is offset by 
0.03 arcseconds to the right compared to the true position; this partly 
accounts for the larger $\delta\theta_c$ in the best-fit model for Simulation 
3, since the subhalo is further from the critical curve.  This shifting of the 
subhalo away from the actual position of the critical curve is a compromise to 
avoid an overly ``sharp'' pinching of the critical curve (and corresponding 
high magnification at the position of the subhalo) that results from the 
steeper density profile in the model subhalo.  Likewise, the size of the 
critical curve perturbation itself may differ slightly as a compromise in order 
to better fit the perturbed surface brightness at different radii; given that 
the surface brightness variations will differ from lens to lens, the bias due 
to this is very difficult to quantify in general.

The second reason for the bias is due to differences in the 
smooth lens component. As is evident in Table \ref{tab:bestfit}, having a 
different log-slope $\alpha$ is not the only difference in the host galaxy 
parameters compared to their actual values.  There is also a slightly different 
external shear $\Gamma_{ext}$ (specifically, the $x$-component $\Gamma_1$) and 
axis ratio $q$, and additionally the host galaxy is shifted significantly to 
the left. All of these change the left-hand side of Eq.~\ref{kapav_eq}, and 
hence the subhalo mass enclosed within $\delta\theta_{c}$ is not strictly 
invariant. Part of this can be specifically accounted for: in Appendix 
\ref{sec:appendix_b} it is shown that, assuming the subhalo position is 
well-reproduced, the invariant quantity is actually 
$m_{sub}(\delta\theta_c)/\alpha\eta = \tilde m_{sub}/\eta$ 
(Eq.~\ref{invariant_mass_eq}), where the parameter $\eta$ is determined by the 
external shear $\Gamma_{ext}$ (Eq.~\ref{eta_on_cc_iso}). Thus, we should have 
$\tilde m_{sub,fit} \approx m_{sub,true}\frac{\eta_{fit}}{\eta_{true}}$. In 
this case the direction of the subhalo $\phi$ and the direction of the external 
shear perturber $\phi_p$ differ by 90 degrees, so we have $\eta \approx 1 + 
\Gamma_{ext} \approx 1 + \Gamma_1$. Thus, using the best-fit values from Table 
\ref{tab:bestfit}, we have $\eta_{true} \approx 1.036$, while $\eta_{fit} 
\approx 1.07$ for Simulations 2-4. Thus, $\tilde m_{sub}$ is inflated by a 
factor of $1.07/1.036 \approx 1.03$, or about 3\% in each case.  While this 
doesn't account for all of the error in $\tilde m_{sub}$, we can see in the 
final column of Table \ref{tab:mass_comparison} that it accounts for roughly 
half the error in Simulations 2 and 4 (though only a quarter of the error in 
Simulation 3; much of the remaining error is likely due to the offset position 
of the subhalo). Unfortunately, the factor $\eta$ doesn't allow us to improve 
our mass estimator, because there is no way to know the true external shear 
(and hence $\eta_{true}$), but it at least allows us to quantify the error due 
to this effect in simulated cases.

Even if it were possible to have prior knowledge of the true parameters of the 
smooth lens component, fixing these to their correct values does not improve 
the fit; indeed, it makes it worse.  The more freedom the smooth lens component 
has to partially mimic the effect of the subhalo beyond $\delta\theta_{c}$, the 
better the subhalo model will be able to reproduce the subhalo perturbation 
radius and hence, the (scaled) mass within $\delta\theta_{c}$. Thus, having 
sufficient freedom to vary the smooth lens component is essential to achieving 
a good fit, even if a small amount of bias inevitably results in these 
parameters as a result of the mismatched subhalo density profile. In fact, it 
may be beneficial to include additional components to the smooth lens model 
beyond what we have done here---in particular, higher-order multipole moments 
might allow for a better fit in the region of the subhalo, but we have not 
explored this option in this work.

\begin{figure*}[t]
	\centering
	\subfigure[variable subhalo mass]
	{
		\includegraphics[height=0.37\hsize,width=0.48\hsize]{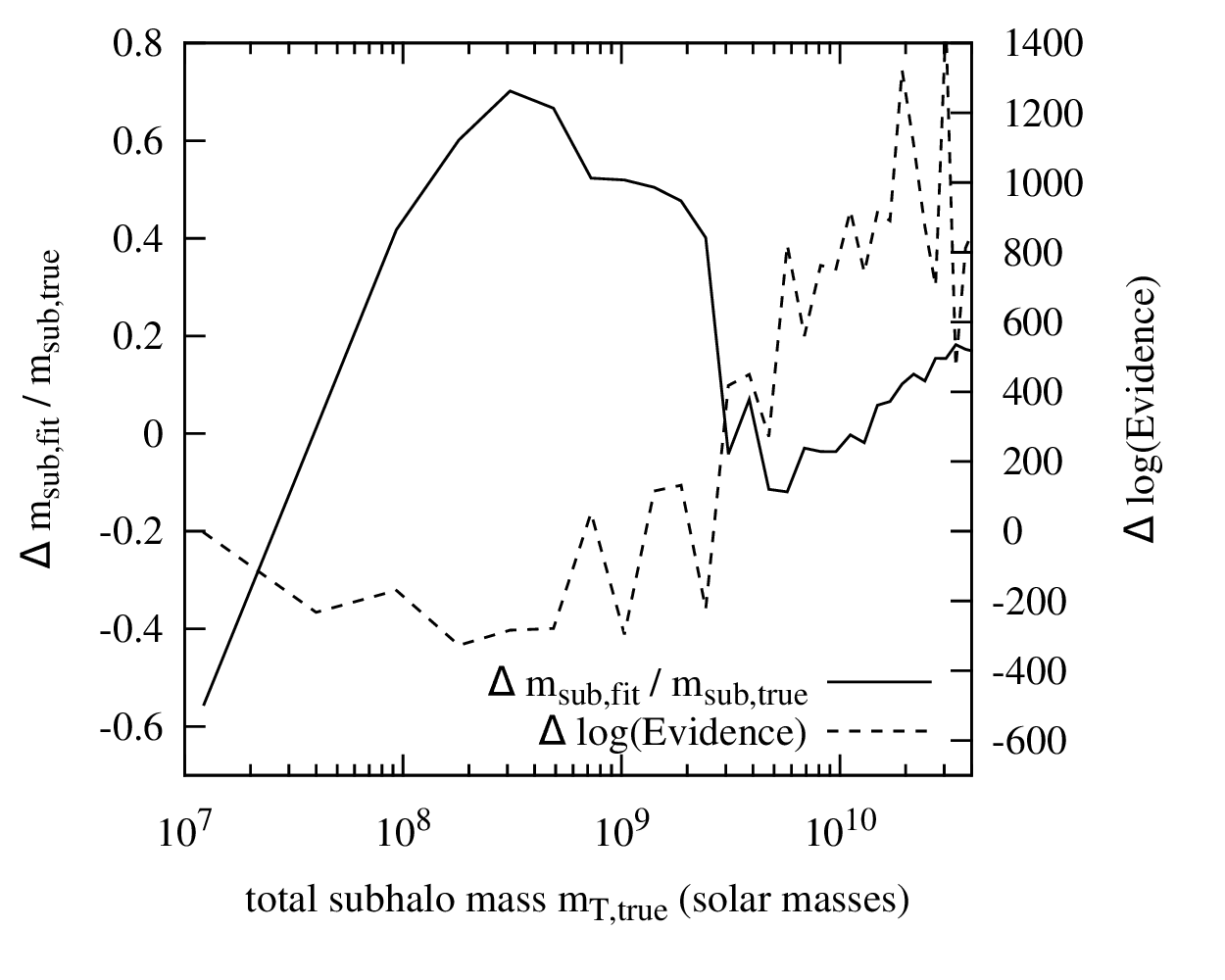}
		\label{fig:mt_mfrac}
	}
	\subfigure[variable subhalo position]
	{
		\includegraphics[height=0.37\hsize,width=0.48\hsize]{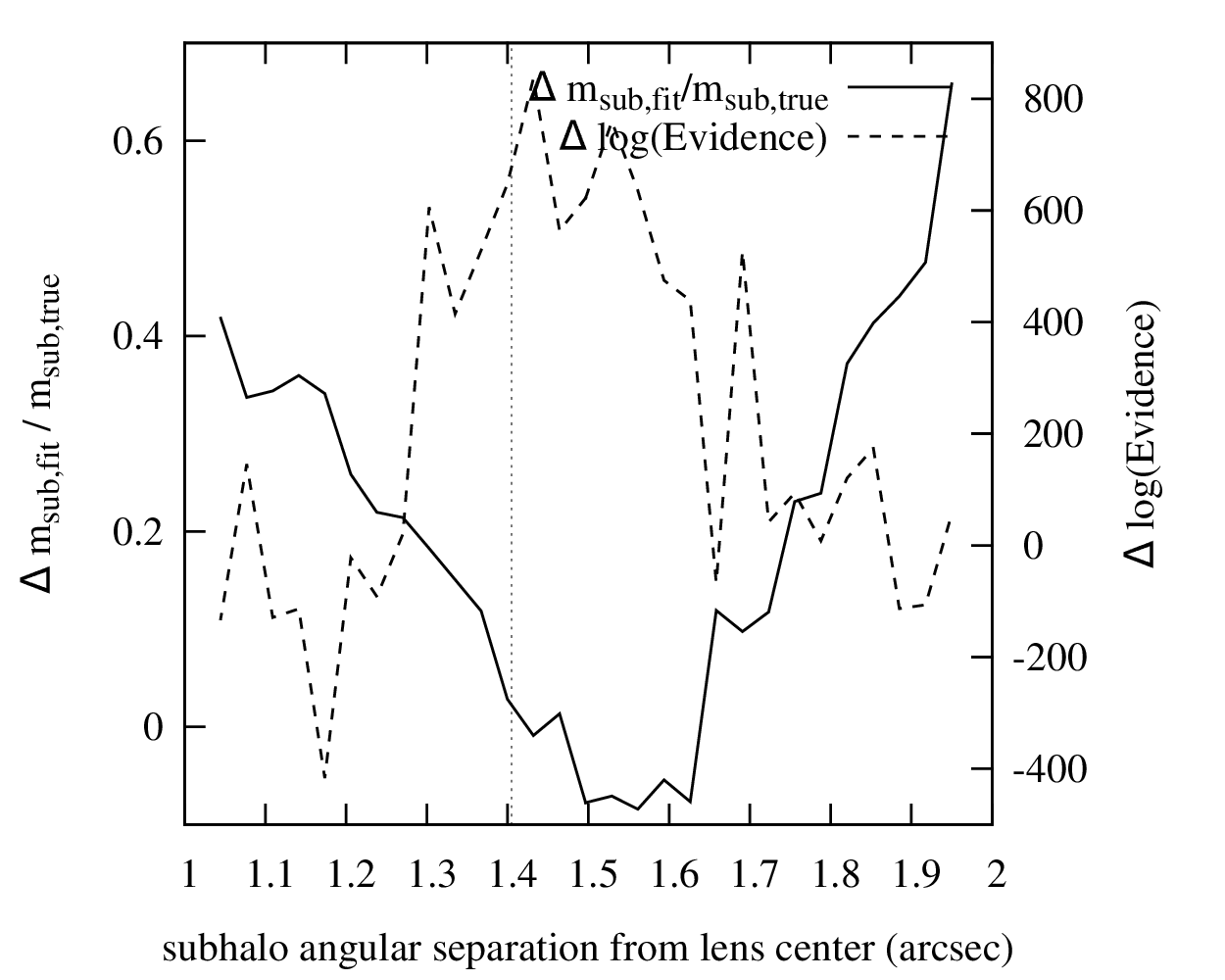}
		\label{fig:rs_mfrac}
	}
	\caption{Fractional error in effective lensing mass $\tilde m_{sub}$ from 
the best-fit subhalo (solid line), for different values of the ``true'' subhalo 
parameters for a subhalo 100 kpc from the lens plane (as in Simulation 4).  In 
(a) the subhalo mass is varied, while in (b) we vary the subhalo's projected 
separation from the host galaxy center; each time the parameter is varied, the 
evidence is minimized to find the best-fit model.  To judge the quality of the 
fit, we compare the Bayesian evidence obtained from the best-fit model with vs.  
without a subhalo included in the model (dotted line); this is quantified by 
$\Delta\log\mathcal{E} = \log\mathcal{E}_{sub} - \log\mathcal{E}_{no sub}$.  
The vertical dotted line in (b) denotes the position of the critical curve 
along the direction of the subhalo. Note that in cases where the effective 
lensing mass is inaccurate by more than $\sim$ 20\%, the Bayesian evidence does 
not prefer including a subhalo to high significance (i.e., 
$\Delta\log\mathcal{E}$ is either negative or is not significantly greater than 
the noise).}
	\label{fig:mtrs_mfrac}
	\vspace{14pt}
\end{figure*}

\section{Range of validity of the effective subhalo lensing 
mass}\label{sec:range}

Thus far we have only tested our subhalo mass estimator $\tilde m_{sub}$ in one 
example lens scenario, in which we have varied only the subhalo density profile 
and/or tidal radius. This begs the question, how accurate is $\tilde m_{sub}$ 
for subhalos of different masses, or at different positions relative to the 
images or critical curve? In this section we will consider different scenarios 
where these subhalo parameters are varied. For simplicity, we will consider
only a subhalo at 100 kpc from the lens plane (as in Simulation 4; see Table 
\ref{tab:sims}), which has a considerably larger tidal radius compared to the 
subhalo model being used in the analysis.  It would be computationally very 
expensive to perform an MCMC over a large number of scenarios, so here we take 
a different approach, by minimizing the Bayesian evidence for each scenario 
using our fiducial subhalo model.

The procedure is as follows. We start with the ``true'' subhalo parameters 
identical to Simulation 4, and minimize the Bayesian evidence to find the 
best-fit model (using our solution in Table~\ref{tab:bestfit} as the initial 
guess parameters). We then vary one of the ``true'' subhalo parameters by a 
small amount, generate a new data image (with different random noise added to 
the image each time), and minimize the Bayesian evidence again, using our 
previous best-fit model as the initial guess parameters; then we repeat the 
cycle. In this way, we generate a ``chain'' of solutions, and for each solution 
we calculate the subhalo perturbation radius numerically and hence, the 
effective lensing mass $\tilde m_{sub}$. To test the quality of the fit, we 
also fit a model that does not include a subhalo and compare the Bayesian 
evidence of the best-fit model with vs. without a subhalo.

The results are shown in Figure \ref{fig:mtrs_mfrac}, where in each figure we 
plot the fractional error in $\tilde m_{sub}$ (solid line, left axis) and the 
difference in the Bayesian Evidence (dashed line, right axis). In Figure 
\ref{fig:mt_mfrac}, we have varied the mass normalization of the subhalo (i.e.  
the parameter $b_s$) and plot with respect to the total subhalo mass. In Figure 
\ref{fig:rs_mfrac}, we varied the subhalo's projected distance from the host 
galaxy center, keeping the subhalo's position angle with respect to the 
$x$-axis fixed; the dotted vertical line denotes the position of the critical 
curve along the direction of the subhalo (for comparison, the image being 
perturbed by the subhalo is roughly at $\sim 1.5$ arcseconds).  In each case, 
note that in the regime where the error in $\tilde m_{sub}$ is less than 
$\sim$20\%, the difference in Bayesian evidence $\Delta\log\mathcal{E}$ is 
positive and significantly greater than the fluctuations generated by the 
random noise, indicating that the subhalo model is favored in these cases.  
Conversely, when $\tilde m_{sub}$ rises well above 20\%, 
$\Delta\log\mathcal{E}$ is either negative or is not significantly greater than 
the fluctuations, indicating that the subhalo detection is not favored in these 
cases.  This gives us confidence that the effective subhalo lensing mass is 
relatively unbiased if the subhalo model is favored by the Bayesian evidence.

Note that in Figure \ref{fig:mt_mfrac}, as the true subhalo mass rises above 
$\sim 5\times 10^9 M_\odot$, the effective lensing mass $\tilde m_{sub}$ 
becomes gradually more inaccurate, even though the subhalo model is strongly 
favored. This is because the fit becomes gradually poorer for very strong 
subhalo perturbations, since the lensing effect of the true subhalo is more 
pronounced and more difficult to reproduce with the assumed subhalo model 
(which has a much smaller tidal radius, in this simulation). Hence the 
effective lensing mass becomes more biased for large subhalo masses.

Although Figure \ref{fig:mt_mfrac} suggests that only subhalos with total mass 
greater than $\sim 3\times 10^9 M_\odot$ can be detected with confidence with 
an ALMA-like resolution ($\sim 0.01$ arcseconds), there are ways to lower this 
detection threshold.  First, smaller mass subhalos should be detectable if both 
the subhalo and the image being perturbed are closer to the critical curve.  
This effect is enhanced if the lens itself is more symmetrical, resulting in 
highly magnified arcs where subhalo perturbations are amplified.  In addition, 
the signal-to-noise ratio in our simulations is $\approx 2$, which is 
relatively low; for long exposures where the signal-to-noise is higher, 
detection of smaller subhalo perturbations becomes easier. Thus,
detection of lower mass subhalos is still possible even with the current 
resolution limit.

\section{Discussion}\label{sec:discussion}

\subsection{Can the subhalo position be well-fit in realistic scenarios?}

While we have demonstrated in Section \ref{sec:results} that our subhalo mass 
estimator $\tilde m_{sub}$ can be determined robustly, we have seen in Section 
\ref{sec:range} that this condition depends on achieving a good fit, such that 
the subhalo model is favored by the Bayesian evidence. In particular, the fit 
must be good enough so that the inferred position of the subhalo is reasonably 
accurate (such that the error in the subhalo's position is small compared to 
the subhalo's distance to the critical curve, $\delta\theta_c$).  One can see 
from Table \ref{tab:bestfit} that this is true for each of the simulations 
analyzed in this paper. However, in Simulation 3 the subhalo's position has the 
greatest error, with the best-fit position shifted by nearly 0.03 arcseconds 
further from the critical curve.  This is a consequence of the true subhalo's 
density profile having a much shallower slope ($\rho \sim r^{-1}$) compared to 
the model subhalo ($\rho \sim r^{-2}$). If the model subhalo is too close to 
the critical curve, it ``pinches'' the critical curve too much (recall the 
difference in perturbation shape from Figure \ref{fig:zoom_imgs}), resulting in 
a perturbation that cannot fit the data well, so a compromise is reached where 
the best-fit subhalo is a bit further out. Indeed, by comparing the Bayesian 
evidence $\mathcal{E}$ in the last column of Table \ref{tab:bestfit}, one sees 
that of the four simulations we modeled, Simulation 3 gave the poorest fit, and 
also gave the greatest error in the effective lensing mass (Table 
\ref{tab:mass_comparison}).  

Is it possible to have a realistic scenario where 
the density profile and tidal radius of the subhalo differ so dramatically from 
the model that a good fit cannot be achieved \emph{at all}?
While we cannot say for sure yet, it is conceivable that a typical NFW subhalo 
may fall in this category. In this paper we varied the suhalo's log-slope and 
tidal radius separately, to isolate each effect. However, if the true subhalo 
has an NFW profile, it can be expected to exhibit a combination of these two 
effects---that is, a shallow inner log-slope of -1, plus a larger tidal radius 
since we expect $r_0 > b$. In this case, the discrepancy in the inferred total 
subhalo mass will be even more profound than it was for Simulation 4, since the 
density profile would not fall off as quickly beyond $\delta\theta_{c}$ and 
more mass would lie outside the subhalo perturbation radius.  More worringly, 
the perturbation of the neighboring images differs so much compared to the 
Pseudo-Jaffe model that it is questionable whether such a scenario could even 
be well-fit by a Pseudo-Jaffe subhalo at all. If so, the implication would be 
that the substructures detected thus far in gravitational lenses \emph{cannot} 
correspond to NFW subhalos with realistic tidal radii.  However, as we have not 
attempted to fit simulated data with an NFW subhalo in this paper, this remains 
an open question at present.

\subsection{Implications of the bias in the inferred density profile of the 
host galaxy}

The effect we have seen on the log-slope of the host galaxy $\alpha$ is rather 
striking, and it is tempting to interpret an $\alpha > 1$ result from an actual 
dataset as evidence that the subhalo density profile is indeed shallower than 
isothermal or that the tidal radius is underestimated.  In \cite{vegetti2010}, 
the inferred log-slope $\alpha \approx 1.2$, which could be consistent with the 
scenarios discussed here. On the other hand, in Vegetti (2012) they infer 
$\alpha \approx 1.05$, while for SDP.81, \cite{hezaveh2016} find $\alpha 
\approx 1.06$; in both cases, the slope of the host galaxy's profile is just 
barely consistent with isothermal.  If taken at face value, this might suggest 
that the subhalo density profile cannot be much shallower than isothermal.  
However, we must be cautious when interpreting these results. It may be the 
case that the host galaxy's density profile may have a log-slope somewhat less 
than 1, in which case these results could still be consistent with a shallower 
subhalo profile.  Indeed, \cite{schneider2013} show that the log-slope of the 
host galaxy in the vicinity of the images can be expected to be biased due to 
the mass-sheet degeneracy if the galaxy's profile differs from a strict 
power-law at other radii.  Furthermore, even if the true $\alpha$ is very close 
to 1 in these cases, the parameter-space solution may not necessarily be 
unique.  As is clear in Figure \ref{fig:triangle_plot}, $\alpha=1$ may still be 
a viable region of parameter space; if $\alpha$ is fixed to 1, the subhalo may 
still be found in that parameter space even if the fit is poorer compared to 
having a steeper $\alpha$. Thus, it is possible that in these cases, a mode 
with higher $\alpha$ might be found by a more complete multimodal sampling of 
the parameter space, e.g. using the T-Walk algorithm, and would produce an even 
better fit.

Ultimately, to minimize the bias in \emph{both} the inferred subhalo properties 
and the host galaxy parameters, the ideal approach is to vary both the 
subhalo's tidal radius $r_t$ and inner log-slope $\gamma$ as free parameters.  
However, while it remains to be seen how well a subhalo's density profile can 
be constrained, e.g. in high-resolution ALMA images of submillimeter 
gravitational lenses, it seems likely that the tidal radius will be 
prior-dominated given that the subhalo pertubation beyond $\delta\theta_{c}$ 
can be mimicked by perturbing the smooth lens component, and hence the exact 
tidal radius will be difficult to constrain. Therefore, given the difficulty in 
constraining these parameters (together with the added computational expense of 
varying them), we argue the effective subhalo lensing mass 
$m_{sub}(\delta\theta_{c})/\alpha$ is a very useful quantity given that it is 
largely robust to changes in the subhalo's density profile and tidal radius.  
Given several subhalo detections, one could use these mass measurements to test 
whether they are consistent with the expected mass function from $\Lambda$CDM, 
or whether an alternate dark matter scenario (e.g., warm dark matter) would 
provide a better fit.

\subsection{Procedure for calculating the subhalo perturbation radius}
\label{sec:tips}

In this section, we provide a summary of procedures for calculating the subhalo 
perturbation radius $\delta\theta_c$ for a given lens model, at different 
levels of approximation.

If the model subhalo is isothermal/Pseudo-Jaffe and one is content with 
approximating $\delta\theta_c$ to 20-30\% accuracy, we recommend using 
Eq.~\ref{rmax_approx_eq2} due to its relative ease of use. Despite the 
approximation, the subhalo mass within this radius is still quite robust (and 
in fact works slightly \emph{better} than the exact values in our Simulations 
2-4). For lens modelers who wish to calculate $\delta\theta_c$ to better 
accuracy, the more rigorous Eq.~\ref{rmax_general} can be used, or 
Eq.~\ref{kapav_eq} can be solved numerically.

To solve for $\delta\theta_c$ numerically using Eq.~\ref{kapav_eq} (which works 
regardless of the model being used for the subhalo profile), one could adopt the 
following method. Calculate the direction $\phi_\Gamma$ of the 
\emph{unperturbed} shear $\boldsymbol{\Gamma_{tot}}$ at the position of the 
subhalo. The solution to Eq.~\ref{kapav_eq} will lie somewhere along the line 
perpendicular to this direction, i.e. along $\phi_r = \phi_\Gamma + 90\degree$, 
because along this line the subhalo's shear aligns with the shear of the 
unperturbed lens model.  In this way, Eq.~\ref{kapav_eq} is reduced to an 
equation in $r$, which can be found with a root-finding algorithm. 

To get a sense of the relative accuracy of these different methods, in Table 
\ref{tab:dtheta_c_comparison} in Appendix \ref{sec:appendix_b} we list a 
comparison of the $\delta\theta_c$ value calculated numerically using this 
procedure, and the values calculated using both Eq.~\ref{rmax_approx_eq2} and 
the more exact Eq.~\ref{rmax_general} for the four simulations considered in 
this paper.

\section{Conclusions}\label{sec:conclusions}

In this article, we have investigated the bias that results from modeling a 
dark matter subhalo in a strong gravitational lens with an incorrect density 
profile. We have derived formulas for the radius of the critical curve 
perturbation generated by subhalos (see Section \ref{sec:tips} for a summary), 
and have shown this scale to be approximately independent of the mass 
distribution of the subhalo within that radius.  We demonstrated the usefulness 
of this scale by modeling a subhalo in a gravitational lens which has a 
different log-slope of the density profile, and different tidal radius, 
compared to the model. Our key results are encapsulated in Figure 
\ref{fig:mprofiles} and Table \ref{tab:mass_comparison}. 

We have found that the estimated total masses of subhalos detected in strong 
gravitational lens systems may be significantly biased because of our ignorance 
about the density profile (tidal radius and inner log-slope) of the perturbing 
subhalo. The most severe error, due to the unknown tidal radius, arises from 
our ignorance of the subhalo's projected distance to the lens plane (this is 
also pointed out by \citealt{vegetti2014b}). However, regardless of this 
ignorance, we show that the effective lensing mass of the perturber, 
$m_{sub}/\alpha$, can be inferred accurately.  The mass $m_{sub}$ is computed 
within a radius $\delta\theta_{c}$, defined by the perturbation of the critical 
curve in the vicinity of the subhalo, while $\alpha$ is the log-slope of the 
density profile of the host.

While we have focused on subhalos in this paper, we note that our mass 
estimator is equally applicable to perturbers along the line-of-sight to the 
lens, unassociated with the primary lens galaxy. In our notation this 
corresponds to the limit $r_0 \rightarrow \infty$, where $r_0$ is the 
perturber's distance from the lens galaxy center.

The effective lensing mass should allow for a more robust comparison to 
simulations and inference of the subhalo mass function from dark matter 
substructure detections. Whether we can go beyond this effective lensing mass 
and gain information about the density profile of a perturber from its effects 
on highly magnified images is an open question that has great relevance
for the nature of dark matter. 

\section*{Acknowledgements}

We thank Simona Vegetti and the anonymous referree for their helpful comments 
on the initial draft.  We also thank Greg Martinez for providing the code for 
the T-Walk algorithm, which proved invaluable for the calculations in this 
article.  QM was supported by NSF grant AST-1615306. MK was supported by NSF 
grant PHY-1316792. NL's work was supported by Argonne National Laboratory under 
the U.S. Department of Energy contract DE-AC02-06CH11357.

We gratefully acknowledge a grant of computer time from XSEDE allocation 
TG-AST130007.

This research was also supported, in part, by a grant of computer time from the 
City University of New York High Performance Computing Center under NSF Grants 
CNS-0855217, CNS-0958379 and ACI-1126113.

\bibliography{subhalo}

\begin{thebibliography}{61}
\expandafter\ifx\csname natexlab\endcsname\relax\def\natexlab#1{#1}\fi

\bibitem[{{Abazajian}(2006)}]{abazajian2006}
{Abazajian}, K. 2006, \prd, 73, 063513

\bibitem[{{ALMA Partnership} {et~al.}(2015){ALMA Partnership}, {Vlahakis},
  {Hunter}, {Hodge}, \& et~al.}]{alma2015}
{ALMA Partnership}, {Vlahakis}, C., {Hunter}, T.~R., {Hodge}, J.~A., \& et~al.
  2015, \apjl, 808, L4

\bibitem[{{Baltz} {et~al.}(2009){Baltz}, {Marshall}, \& {Oguri}}]{baltz2009}
{Baltz}, E.~A., {Marshall}, P., \& {Oguri}, M. 2009, \jcap, 1, 015

\bibitem[{{Barkana}(1998)}]{barkana1998}
{Barkana}, R. 1998, \apj, 502, 531

\bibitem[{{Bode} {et~al.}(2001){Bode}, {Ostriker}, \& {Turok}}]{bode2001}
{Bode}, P., {Ostriker}, J.~P., \& {Turok}, N. 2001, \apj, 556, 93

\bibitem[{{Boylan-Kolchin} {et~al.}(2011){Boylan-Kolchin}, {Bullock}, \&
  {Kaplinghat}}]{boylan2011}
{Boylan-Kolchin}, M., {Bullock}, J.~S., \& {Kaplinghat}, M. 2011, \mnras, 415,
  L40

\bibitem[{{Boylan-Kolchin} {et~al.}(2012){Boylan-Kolchin}, {Bullock}, \&
  {Kaplinghat}}]{boylan2012}
---. 2012, \mnras, 422, 1203

\bibitem[{{Budzynski} {et~al.}(2012){Budzynski}, {Koposov}, {McCarthy},
  {McGee}, \& {Belokurov}}]{budzynski2012}
{Budzynski}, J.~M., {Koposov}, S.~E., {McCarthy}, I.~G., {McGee}, S.~L., \&
  {Belokurov}, V. 2012, \mnras, 423, 104

\bibitem[{{Bullock} {et~al.}(2000){Bullock}, {Kravtsov}, \&
  {Weinberg}}]{bullock2000}
{Bullock}, J.~S., {Kravtsov}, A.~V., \& {Weinberg}, D.~H. 2000, \apj, 539, 517

\bibitem[{{Chen}(2009)}]{chen2009}
{Chen}, J. 2009, \aap, 494, 867

\bibitem[{Christen \& Fox(2010)}]{christen2010}
Christen, J.~A. \& Fox, C. 2010, Bayesian Anal., 5, 263

\bibitem[{{Cyr-Racine} {et~al.}(2016){Cyr-Racine}, {Moustakas}, {Keeton},
  {Sigurdson}, \& {Gilman}}]{cyrracine2016}
{Cyr-Racine}, F.-Y., {Moustakas}, L.~A., {Keeton}, C.~R., {Sigurdson}, K., \&
  {Gilman}, D.~A. 2016, \prd, 94, 043505

\bibitem[{{Dalal} \& {Kochanek}(2002)}]{dalal2002}
{Dalal}, N. \& {Kochanek}, C.~S. 2002, \apj, 572, 25

\bibitem[{{Despali} \& {Vegetti}(2016)}]{despali2016}
{Despali}, G. \& {Vegetti}, S. 2016, ArXiv e-prints

\bibitem[{{Diemand} {et~al.}(2008){Diemand}, {Kuhlen}, {Madau}, {Zemp},
  {Moore}, {Potter}, \& {Stadel}}]{diemand2008}
{Diemand}, J., {Kuhlen}, M., {Madau}, P., {Zemp}, M., {Moore}, B., {Potter},
  D., \& {Stadel}, J. 2008, \nat, 454, 735

\bibitem[{{Dye} {et~al.}(2015){Dye}, {Furlanetto}, {Swinbank}, {Vlahakis},
  {Nightingale}, {Dunne}, {Eales}, {Smail}, {Oteo}, {Hunter}, {Negrello},
  {Dannerbauer}, {Ivison}, {Gavazzi}, {Cooray}, \& {van der Werf}}]{dye2015}
{Dye}, S., {Furlanetto}, C., {Swinbank}, A.~M., {Vlahakis}, C., {Nightingale},
  J.~W., {Dunne}, L., {Eales}, S.~A., {Smail}, I., {Oteo}, I., {Hunter}, T.,
  {Negrello}, M., {Dannerbauer}, H., {Ivison}, R.~J., {Gavazzi}, R., {Cooray},
  A., \& {van der Werf}, P. 2015, \mnras, 452, 2258

\bibitem[{{Dye} \& {Warren}(2005)}]{dye2005}
{Dye}, S. \& {Warren}, S.~J. 2005, \apj, 623, 31

\bibitem[{{Elbert} {et~al.}(2014){Elbert}, {Bullock}, {Garrison-Kimmel},
  {Rocha}, {O{\~n}orbe}, \& {Peter}}]{elbert2014}
{Elbert}, O.~D., {Bullock}, J.~S., {Garrison-Kimmel}, S., {Rocha}, M.,
  {O{\~n}orbe}, J., \& {Peter}, A.~H.~G. 2014, ArXiv e-prints

\bibitem[{{Flores} \& {Primack}(1994)}]{flores1994}
{Flores}, R.~A. \& {Primack}, J.~R. 1994, \apjl, 427, L1

\bibitem[{{Governato} {et~al.}(2012){Governato}, {Zolotov}, {Pontzen},
  {Christensen}, {Oh}, {Brooks}, {Quinn}, {Shen}, \& {Wadsley}}]{governato2012}
{Governato}, F., {Zolotov}, A., {Pontzen}, A., {Christensen}, C., {Oh}, S.~H.,
  {Brooks}, A.~M., {Quinn}, T., {Shen}, S., \& {Wadsley}, J. 2012, \mnras, 422,
  1231

\bibitem[{{Guo} {et~al.}(2012){Guo}, {Cole}, {Eke}, \& {Frenk}}]{guo2012}
{Guo}, Q., {Cole}, S., {Eke}, V., \& {Frenk}, C. 2012, \mnras, 427, 428

\bibitem[{{Hansen} {et~al.}(2005){Hansen}, {McKay}, {Wechsler}, {Annis},
  {Sheldon}, \& {Kimball}}]{hansen2005}
{Hansen}, S.~M., {McKay}, T.~A., {Wechsler}, R.~H., {Annis}, J., {Sheldon},
  E.~S., \& {Kimball}, A. 2005, \apj, 633, 122

\bibitem[{{Hatsukade} {et~al.}(2015){Hatsukade}, {Tamura}, {Iono}, {Matsuda},
  {Hayashi}, \& {Oguri}}]{hatsukade2015}
{Hatsukade}, B., {Tamura}, Y., {Iono}, D., {Matsuda}, Y., {Hayashi}, M., \&
  {Oguri}, M. 2015, \pasj, 67, 93

\bibitem[{{Hezaveh} {et~al.}(2016){Hezaveh}, {Dalal}, {Marrone}, {Mao},
  {Morningstar}, {Wen}, {Blandford}, {Carlstrom}, {Fassnacht}, {Holder},
  {Kemball}, {Marshall}, {Murray}, {Perreault Levasseur}, {Vieira}, \&
  {Wechsler}}]{hezaveh2016}
{Hezaveh}, Y.~D., {Dalal}, N., {Marrone}, D.~P., {Mao}, Y.-Y., {Morningstar},
  W., {Wen}, D., {Blandford}, R.~D., {Carlstrom}, J.~E., {Fassnacht}, C.~D.,
  {Holder}, G.~P., {Kemball}, A., {Marshall}, P.~J., {Murray}, N., {Perreault
  Levasseur}, L., {Vieira}, J.~D., \& {Wechsler}, R.~H. 2016, \apj, 823, 37

\bibitem[{{Inoue} {et~al.}(2016){Inoue}, {Minezaki}, {Matsushita}, \&
  {Chiba}}]{inoue2016}
{Inoue}, K.~T., {Minezaki}, T., {Matsushita}, S., \& {Chiba}, M. 2016, \mnras,
  457, 2936

\bibitem[{{Keeton} \& {Moustakas}(2009)}]{keeton2009}
{Keeton}, C.~R. \& {Moustakas}, L.~A. 2009, \apj, 699, 1720

\bibitem[{{Klypin} {et~al.}(1999){Klypin}, {Kravtsov}, {Valenzuela}, \&
  {Prada}}]{klypin1999}
{Klypin}, A., {Kravtsov}, A.~V., {Valenzuela}, O., \& {Prada}, F. 1999, \apj,
  522, 82

\bibitem[{{Kochanek} \& {Dalal}(2004)}]{kochanek2004}
{Kochanek}, C.~S. \& {Dalal}, N. 2004, \apj, 610, 69

\bibitem[{{Koopmans}(2005)}]{koopmans2005}
{Koopmans}, L.~V.~E. 2005, \mnras, 363, 1136

\bibitem[{{Kuzio de Naray} {et~al.}(2006){Kuzio de Naray}, {McGaugh}, {de
  Blok}, \& {Bosma}}]{kuzio2006}
{Kuzio de Naray}, R., {McGaugh}, S.~S., {de Blok}, W.~J.~G., \& {Bosma}, A.
  2006, \apjs, 165, 461

\bibitem[{{Lovell} {et~al.}(2014){Lovell}, {Frenk}, {Eke}, {Jenkins}, {Gao}, \&
  {Theuns}}]{lovell2014}
{Lovell}, M.~R., {Frenk}, C.~S., {Eke}, V.~R., {Jenkins}, A., {Gao}, L., \&
  {Theuns}, T. 2014, \mnras, 439, 300

\bibitem[{{Mao} \& {Schneider}(1998)}]{mao1998}
{Mao}, S. \& {Schneider}, P. 1998, \mnras, 295, 587

\bibitem[{{Metcalf} \& {Madau}(2001)}]{metcalf2001}
{Metcalf}, R.~B. \& {Madau}, P. 2001, \apj, 563, 9

\bibitem[{{Moore}(1994)}]{moore1994}
{Moore}, B. 1994, \nat, 370, 629

\bibitem[{{Moustakas} \& {Metcalf}(2003)}]{moustakas2003}
{Moustakas}, L.~A. \& {Metcalf}, R.~B. 2003, \mnras, 339, 607

\bibitem[{{Mu{\~n}oz} {et~al.}(2001){Mu{\~n}oz}, {Kochanek}, \&
  {Keeton}}]{munoz2001}
{Mu{\~n}oz}, J.~A., {Kochanek}, C.~S., \& {Keeton}, C.~R. 2001, \apj, 558, 657

\bibitem[{{Navarro} {et~al.}(1996){Navarro}, {Frenk}, \& {White}}]{navarro1996}
{Navarro}, J.~F., {Frenk}, C.~S., \& {White}, S.~D.~M. 1996, \apj, 462, 563

\bibitem[{{Navarro} {et~al.}(2010){Navarro}, {Ludlow}, {Springel}, {Wang},
  {Vogelsberger}, {White}, {Jenkins}, {Frenk}, \& {Helmi}}]{navarro2010}
{Navarro}, J.~F., {Ludlow}, A., {Springel}, V., {Wang}, J., {Vogelsberger}, M.,
  {White}, S.~D.~M., {Jenkins}, A., {Frenk}, C.~S., \& {Helmi}, A. 2010,
  \mnras, 402, 21

\bibitem[{{Nierenberg} {et~al.}(2012){Nierenberg}, {Auger}, {Treu}, {Marshall},
  {Fassnacht}, \& {Busha}}]{nierenberg2012}
{Nierenberg}, A.~M., {Auger}, M.~W., {Treu}, T., {Marshall}, P.~J.,
  {Fassnacht}, C.~D., \& {Busha}, M.~T. 2012, \apj, 752, 99

\bibitem[{{Nierenberg} {et~al.}(2014){Nierenberg}, {Treu}, {Wright},
  {Fassnacht}, \& {Auger}}]{nierenberg2014}
{Nierenberg}, A.~M., {Treu}, T., {Wright}, S.~A., {Fassnacht}, C.~D., \&
  {Auger}, M.~W. 2014, \mnras, 442, 2434

\bibitem[{{Nightingale} \& {Dye}(2015)}]{nightingale2015}
{Nightingale}, J.~W. \& {Dye}, S. 2015, \mnras, 452, 2940

\bibitem[{{Rocha} {et~al.}(2013){Rocha}, {Peter}, {Bullock}, {Kaplinghat},
  {Garrison-Kimmel}, {O{\~n}orbe}, \& {Moustakas}}]{rocha2013}
{Rocha}, M., {Peter}, A.~H.~G., {Bullock}, J.~S., {Kaplinghat}, M.,
  {Garrison-Kimmel}, S., {O{\~n}orbe}, J., \& {Moustakas}, L.~A. 2013, \mnras,
  430, 81

\bibitem[{{Rybak} {et~al.}(2015){Rybak}, {McKean}, {Vegetti}, {Andreani}, \&
  {White}}]{rybak2015}
{Rybak}, M., {McKean}, J.~P., {Vegetti}, S., {Andreani}, P., \& {White},
  S.~D.~M. 2015, \mnras, 451, L40

\bibitem[{{Schneider} \& {Sluse}(2013)}]{schneider2013}
{Schneider}, P. \& {Sluse}, D. 2013, \aap, 559, A37

\bibitem[{{Spergel} \& {Steinhardt}(2000)}]{spergel2000}
{Spergel}, D.~N. \& {Steinhardt}, P.~J. 2000, Physical Review Letters, 84, 3760

\bibitem[{{Springel} {et~al.}(2008){Springel}, {Wang}, {Vogelsberger},
  {Ludlow}, {Jenkins}, {Helmi}, {Navarro}, {Frenk}, \& {White}}]{springel2008}
{Springel}, V., {Wang}, J., {Vogelsberger}, M., {Ludlow}, A., {Jenkins}, A.,
  {Helmi}, A., {Navarro}, J.~F., {Frenk}, C.~S., \& {White}, S.~D.~M. 2008,
  \mnras, 391, 1685

\bibitem[{{Suyu} \& {Halkola}(2010)}]{suyu2010}
{Suyu}, S.~H. \& {Halkola}, A. 2010, \aap, 524, A94

\bibitem[{{Suyu} {et~al.}(2006){Suyu}, {Marshall}, {Hobson}, \&
  {Blandford}}]{suyu2006}
{Suyu}, S.~H., {Marshall}, P.~J., {Hobson}, M.~P., \& {Blandford}, R.~D. 2006,
  \mnras, 371, 983

\bibitem[{{Tagore} \& {Keeton}(2014)}]{tagore2014}
{Tagore}, A.~S. \& {Keeton}, C.~R. 2014, \mnras, 445, 694

\bibitem[{{Tamura} {et~al.}(2015){Tamura}, {Oguri}, {Iono}, {Hatsukade},
  {Matsuda}, \& {Hayashi}}]{tamura2015}
{Tamura}, Y., {Oguri}, M., {Iono}, D., {Hatsukade}, B., {Matsuda}, Y., \&
  {Hayashi}, M. 2015, \pasj, 67, 72

\bibitem[{{Tessore} \& {Metcalf}(2015)}]{tessore2015}
{Tessore}, N. \& {Metcalf}, R.~B. 2015, \aap, 580, A79

\bibitem[{{Vegetti} \& {Koopmans}(2009)}]{vegetti2009}
{Vegetti}, S. \& {Koopmans}, L.~V.~E. 2009, \mnras, 392, 945

\bibitem[{{Vegetti} {et~al.}(2014){Vegetti}, {Koopmans}, {Auger}, {Treu}, \&
  {Bolton}}]{vegetti2014b}
{Vegetti}, S., {Koopmans}, L.~V.~E., {Auger}, M.~W., {Treu}, T., \& {Bolton},
  A.~S. 2014, \mnras, 442, 2017

\bibitem[{{Vegetti} {et~al.}(2010){Vegetti}, {Koopmans}, {Bolton}, {Treu}, \&
  {Gavazzi}}]{vegetti2010}
{Vegetti}, S., {Koopmans}, L.~V.~E., {Bolton}, A., {Treu}, T., \& {Gavazzi}, R.
  2010, \mnras, 408, 1969

\bibitem[{{Vegetti} {et~al.}(2012){Vegetti}, {Lagattuta}, {McKean}, {Auger},
  {Fassnacht}, \& {Koopmans}}]{vegetti2012}
{Vegetti}, S., {Lagattuta}, D.~J., {McKean}, J.~P., {Auger}, M.~W.,
  {Fassnacht}, C.~D., \& {Koopmans}, L.~V.~E. 2012, \nat, 481, 341

\bibitem[{{Vegetti} \& {Vogelsberger}(2014)}]{vegetti2014}
{Vegetti}, S. \& {Vogelsberger}, M. 2014, \mnras, 442, 3598

\bibitem[{{Wang} {et~al.}(2014){Wang}, {Sales}, {Henriques}, \&
  {White}}]{wang2014}
{Wang}, W., {Sales}, L.~V., {Henriques}, B.~M.~B., \& {White}, S.~D.~M. 2014,
  \mnras, 442, 1363

\bibitem[{{Warren} \& {Dye}(2003)}]{warren2003}
{Warren}, S.~J. \& {Dye}, S. 2003, \apj, 590, 673

\bibitem[{{Wojtak} \& {Mamon}(2013)}]{wojtak2013}
{Wojtak}, R. \& {Mamon}, G.~A. 2013, \mnras, 428, 2407

\bibitem[{{Wong} {et~al.}(2015){Wong}, {Suyu}, \& {Matsushita}}]{wong2015}
{Wong}, K.~C., {Suyu}, S.~H., \& {Matsushita}, S. 2015, \apj, 811, 115

\bibitem[{{Xu} {et~al.}(2015){Xu}, {Sluse}, {Gao}, {Wang}, {Frenk}, {Mao},
  {Schneider}, \& {Springel}}]{xu2015}
{Xu}, D., {Sluse}, D., {Gao}, L., {Wang}, J., {Frenk}, C., {Mao}, S.,
  {Schneider}, P., \& {Springel}, V. 2015, \mnras, 447, 3189

\end{thebibliography}

\appendix

\setcounter{section}{0}
\section{Appendix A: Equation for critical curve perturbations by 
subhalos}\label{sec:appendix_a}

In this section we derive the general equation for the size of the perturbation 
to a critical curve generated by a subhalo.

Suppose we have a gravitational lens with a host galaxy whose density profile 
is denoted by $\kappa_0$, an external shear $\Gamma_{ext}$, and subhalo denoted 
by $\kappa_s$. To find the formula for the inverse magnification at a given 
position, we adopt a coordinate system where the combined shear of the host 
galaxy and the external shear is diagonalized. In this coordinate system, the 
Jacobian becomes

\begin{equation}
\frac{\partial{\boldsymbol{\beta}}}{\partial\boldsymbol{\theta}} = \left( \begin{array}{cc} 1 - (\kappa_0 + \kappa_s) - (\Gamma_{tot} - \Gamma_s \cos 2\phi_s) & \Gamma_s \sin 2\phi_s \\ \Gamma_s \sin 2\phi_s & 1 - (\kappa_0 + \kappa_s) + (\Gamma_{tot} - \Gamma_s \cos 2\phi_s) 
\end{array} \right)
\end{equation}
where $\phi_s$ gives the direction of the subhalo relative to the direction of 
the total unperturbed shear, $\boldsymbol{\Gamma_{tot}} = \boldsymbol{\Gamma_0} 
+ \boldsymbol{\Gamma_{ext}}$.  (Note, $\phi_s$ is \emph{not} the shear direction 
of the subhalo, but rather points toward the subhalo itself. In this formula, 
the subhalo is assumed to be axisymmetric, so the subhalo's shear is 
perpendicular to the direction of the subhalo---this accounts for the minus 
sign in front of $\Gamma_s$.) The inverse magnification is then the determinant 
of the Jacobian, which can be written as

\begin{equation}
\mu^{-1} = (\lambda_- - \kappa_s + \Gamma_s \cos 2\phi_s)(\lambda_+ - \kappa_s - \Gamma_s \cos 2\phi_s) - \Gamma_s^2 \sin^2 2\phi_s
\label{muinv_formula}
\end{equation}
where the $\lambda_\pm$ are the eigenvalues of the magnification of the 
unperturbed model,

\begin{equation}
\lambda_{\pm} = 1 - \kappa_0 \pm \Gamma_{tot}.
\end{equation}

The set of points on the lens plane for which $\lambda_- = 0$ corresponds to 
the tangential critical curve, while $\lambda_+ = 0$ corresponds to the radial 
critical curve. Radial critical curves do not exist if the host galaxy is 
isothermal, and since their effects are difficult to observe in any case, we 
will therefore focus on perturbations to the tangential critical curve.

After some algebraic manipulation, this can be cast into the following equation:

\begin{equation}
\mu^{-1} = \mu_0^{-1} + \mu_s^{-1} - 1 + 2\kappa_0\kappa_s + 2\Gamma_{tot}\Gamma_s \cos 2\phi_s
\label{muinv_formula2}
\end{equation}

Inside the tangential critical curve, the inverse magnification $\mu^{-1}$ is 
negative. In the case of an axisymmetric host galaxy with no external 
shear present, we define the direction of maximum warping of the critical curve 
to be the direction along which the critical curve is pushed furthest away from 
the subhalo, compared to its position without the subhalo present.  This will 
be the direction for which the inverse magnification $\mu^{-1}$ is decreased 
the most (or rather, made more negative). According to 
Eq.~\ref{muinv_formula2}, this occurs when $\phi_s=90\degree$, the direction 
along which the shear of the subhalo aligns with that of the host galaxy.

If the host galaxy is axisymmetric, its shear is parallel to the tangential 
critical curve, and thus $\phi_s=90\degree$ would be along the radial direction.  
For a non-axisymmetric host galaxy with external shear, the ``direction of 
maximum warping'' is a bit more ambiguous, because the comparison depends on 
how the critical curve is parameterized with and without the subhalo.  Since 
the ambiguity only arises for highly asymmetric lenses and we are mainly 
interested in the overall scale of the perturbation, we choose the simplest 
approach and define $\phi_s=90\degree$ to be the direction of maximal warping in 
general.

If we return to Eq.~\ref{muinv_formula} and set $\phi_s=90\degree$ and 
$\mu^{-1}=0$ to find the location of the tangential critical curve, we find 

\begin{equation}
\lambda_-(\boldsymbol{\theta}) = \kappa_s + \Gamma_s.
\end{equation}
Although there is also a radial critical curve solution, we are focusing only 
on the tangential critical curve solution involving $\lambda_-$. Using the fact 
that for an axisymmetric lens we have $\Gamma = \bar\kappa - \kappa$, we can 
rewrite the equation as follows,

\begin{equation}
1 - \kappa_0(\boldsymbol{\theta_s}+\boldsymbol{r}) - \Gamma_{tot}(\boldsymbol{\theta_s}+\boldsymbol{r}) = \bar\kappa_s(r)
\label{kapav_eq_app}
\end{equation}
where $\boldsymbol{r} = \boldsymbol{\theta} - \boldsymbol{\theta_s}$, and 
$\boldsymbol{\theta_s}$ is the position of the subhalo. Once the direction of 
$\boldsymbol r$ is fixed by the $\phi_s=90\degree$ requirement, this equation can 
be solved for $r$ to find the radius of maximum deformation of the critical 
curve.  From the formula it is evident that the solution, which we call the ``subhalo perturbation radius'' $\delta\theta_c$, only depends on the 
mass of the subhalo enclosed within the radius $\delta\theta_c$ (and hence, the average 
kappa enclosed); it does not depend on how that mass is distributed within this radius.

\section{Appendix B: General solution for the subhalo perturbation radius and 
effective lensing mass}\label{sec:appendix_b}

\subsection{B1. General analytic equation for the subhalo perturbation radius}

Here we will obtain a more rigorous result for the subhalo perturbation radius, 
in the case where the subhalo is not exactly on a critical curve and both 
ellipticity and external shear are present.  Suppose the subhalo is located at 
a radius $\theta_s$ (not necessarily on the critical curve) and angle $\phi$ 
with respect to the $x$-axis, and the host galaxy has an axis ratio $q$ and 
major axis oriented at an angle $\phi_0$ with respect to the $x$-axis.  Suppose 
also that there is an external shear $\Gamma_{ext}$ created by a distant 
perturbing mass whose direction is denoted by angle $\phi_p$.

The projected density profile of the host galaxy is modeled as a power-law ellipsoid. In a 
coordinate system ($x',y'$) where the major axis of the lens coincides with the 
$x'$-axis, we write the profile as follows, using the notation 
$\boldsymbol{\theta'} = (x',y')$:

\begin{equation}
\kappa_0(\boldsymbol \theta') = \frac{2-\alpha}{2}b^\alpha\left(qx'^2 + y'^2/q\right)^{-\frac{\alpha}{2}}
\label{kappa0_eq_app}
\end{equation}

In this formulation, the normalization parameter $b$ can be thought of as the average Einstein radius, 
while the semimajor axis of the critical curve is $b/\sqrt{q}$.  Another common 
choice is to make the semimajor axis of the critical curve the normalization parameter (call 
it $b'$), in which case one should substitute $b' = b/\sqrt{q}$ in the 
following formulas.  In any event, in the actual coordinate system of the lens 
plane ($x,y$), we calculate $\kappa_0(\boldsymbol\theta)$ by rotating these 
coordinates by the angle $\phi_0$; in other words, by first calculating
$\theta_{i}' = \mathcal{R}_{ij}\theta_j$ where $\mathcal{R}_{ij}$ is the 
corresponding rotation matrix, and inserting into Eq.~\ref{kappa0_eq_app}.

We can write this in terms of polar coordinates in the lens frame $(\theta,\phi)$ and
separate out the angular dependence by rewriting it as follows:

\begin{equation}
\kappa_0(\boldsymbol \theta) = \frac{2-\alpha}{2}\left(\frac{b\sigma}{\theta}\right)^\alpha
\label{kappa0_eq}
\end{equation}
where
\begin{equation}
\sigma \equiv \left(q\cos^2(\phi-\phi_0) + \frac{1}{q}\sin^2(\phi-\phi_0)\right)^{-\frac{1}{2}}.
\end{equation}

To solve Eq.~\ref{kapav_eq_app}, we need an expression for the shear from the 
host galaxy $\Gamma_0$. For an isothermal ellipsoid ($\alpha=1$), the magnitude 
of the shear is equal to $\kappa_0$ at that location, so that $\kappa_0 + 
\Gamma_0 = 2\kappa_0$. For other values of $\alpha$ this is no longer true, but 
it can be shown that the ratio $g \equiv \Gamma_0/\kappa_0$ depends only on the 
angle $\phi-\phi_0$; for a fixed angle, $g$ is the same at all radii. This 
ratio can be found numerically using gravitational lensing software, but it can 
also be derived analytically using the results of \cite{tessore2015}, where a 
closed-form solution for the deflection of a power-law ellipsoid lens is 
derived.  The result is

\begin{equation}
g = \left|1 - \frac{4(1-\alpha)\sqrt{q}}{(1+q)(2-\alpha)\sigma}e^{i(\Delta\phi'-\Delta\phi)} {_2}F_1\left(1;\frac{\alpha}{2};2-\frac{\alpha}{2};\frac{q-1}{q+1}e^{i2\Delta\phi}\right)\right|
\label{g_eq}
\end{equation}
where ${_2}F_{1}(\dots)$ is the complex Gaussian hypergeometric function, and 
$\Delta\phi'$ is defined by $\tan\Delta\phi' = \frac{1}{q}\tan\Delta\phi$, 
where it is understood that $\Delta\phi'$ must be in the same quadrant as 
$\Delta\phi$. Note that in the isothermal case $\alpha=1$, Eq.~\ref{g_eq} 
reduces to $g=1$, as expected.

If the axis ratio $q=1$, Eq.~\ref{g_eq} reduces to the much simpler expression 
$g = \alpha/(2-\alpha)$, which is consistent with the fact that $\Gamma_0 = 
\bar\kappa_0 - \kappa_0$ when the host galaxy is axisymmetric.  This expression 
may be used as a rough approximation for $g$ in the regime $q > 0.8$, $0.8 < 
\alpha < 1.2$ or so with only a few percent loss of accuracy in 
$\delta\theta_c$; for $q$ and $\alpha$ outside this range however, the error 
can become severe, so in that case we recommend using Eq.~\ref{g_eq} or else 
calculating $g$ using gravitational lensing software.

The direction of the shear is also important, since this determines the 
direction of maximal warping of the critical curve. In the $\alpha=1$ case, it 
can be shown algebraically that the shear from an isothermal ellipsoid is still 
along the tangential direction, even for small $q$-values (and even though the 
critical curve is \emph{not} along the tangential direction!).  Thus, in the 
absence of external shear, the direction of maximal warping of the critical 
curve by a subhalo is still along the radial direction for an isothermal galaxy. For $\alpha \neq 1$ 
this is no longer strictly true; however we find that as long as $\alpha > 
0.7$, the difference compared to the radial direction is no more than 
$5\degree$. The presence of external shear can also affect the shear direction, 
however.  At a point given by polar coordinates $(\theta,\phi)$, the total 
shear magnitude and direction are given by

\begin{equation}
\Gamma_{tot} \approx \sqrt{g^2\kappa_0^2+\Gamma_{ext}^2 + 2g\kappa_0\Gamma_{ext}\cos 
2(\phi-\phi_p)},
\end{equation}
\begin{equation}
\phi_\Gamma \approx \frac{\pi}{2} + \frac{1}{2}\sin^{-1}\left(\frac{g\kappa_0\sin 2\phi 
+ \Gamma_{ext}\sin 2\phi_p}{\Gamma_{tot}}\right).
\end{equation}
where we have made the approximation that the shear of the host galaxy is along 
the radial direction (which is exactly true if $\alpha=1$).  It should be 
emphasized that $\phi_p$ is the direction of the (distant) perturbing galaxy 
producing the external shear, \emph{not} the direction of the external shear 
itself (which differs from $\phi_p$ by $90\degree$). Note that for 
$\Gamma_{ext}=0$, the shear angle $\phi_{\Gamma,0} = \phi + \frac{\pi}{2}$, 
i.e. the shear is tangential in that case. However, the difference in the shear 
angle due to the external shear, $\Delta\phi_\Gamma = \phi_\Gamma - 
\phi_{\Gamma,0}$ is fairly small provided that $\Gamma_{ext} \lesssim 0.2$; in 
this case we find that typically $\Delta\phi_\Gamma \lesssim 10\degree$, and 
the corresponding change in $\delta\theta_{c}$ due to this direction change was within a 
few percent. Therefore, in the following derivation we will approximate the 
direction of $\delta\theta_{c}$ to still be along the radial direction, given that the 
effect of the perturbed shear direction is rather small.

To derive the formula for $\delta\theta_{c}$, we return to Eq.~\ref{kapav_eq_app} and 
rewrite it as follows:

\begin{equation}
\eta - (1+g)\kappa_0(\boldsymbol{\theta_s}+\boldsymbol{r}) = \bar\kappa_s(r),
\label{rmax_eq_eta}
\end{equation}
where we have defined
\begin{eqnarray}
\eta & \equiv & 1 + \Gamma_0 - \Gamma_{tot} \nonumber \\
& = & 1 + g\kappa_0 - g\kappa_0\sqrt{1+\left(\frac{\Gamma_{ext}}{g\kappa_0}\right)^2 + 2\frac{\Gamma_{ext}}{g\kappa_0}\cos 2(\phi-\phi_p)}. 
\label{eta_eq}
\end{eqnarray}

As long as the external shear is not too small ($\Gamma_{ext} \lesssim 0.3$ or 
so), $\eta$ varies only a miniscule amount as $r$ is varied (in fact, it is 
easy to show that if we choose the direction $\phi = \phi_p$ or 
$\phi=\phi_p+\frac{\pi}{2}$, it does not vary at all with $r$; only for 
directions unaligned with the external shear is there a slight variation).  
Because it varies so little, a very good approximation can be found by fixing 
$\eta$ to its value at the (unperturbed) critical curve along the direction of 
the subhalo. Using the fact that on the unperturbed critical curve we have 
$1-\kappa_0 - \Gamma_{tot} = 0$, after some tedious algebra, we derive

\begin{eqnarray}
& \eta ~ ~ \approx & \frac{1 + g\Gamma_{ext}\cos 2(\phi-\phi_p)}{g-1} \nonumber \\
& ~ ~ ~ ~ & \times ~ ~ \left\{-1 + \sqrt{1 + \frac{(g^2-1)(1-\Gamma_{ext}^2)}{(1 + g\Gamma_{ext}\cos 2(\phi-\phi_p))^2}}\right\}.
\label{eta_on_cc}
\end{eqnarray}

In general this expression increases extremely slowly with $g$, so a very good 
approximation is achieved by taking the limiting case $g \rightarrow 1$ (which 
corresponds to the isothermal case):

\begin{equation}
\eta \approx \frac{1-\Gamma_{ext}^2}{1+\Gamma_{ext}\cos2(\phi-\phi_p)}.  
\label{eta_on_cc_iso}
\end{equation}

In order for Eq.~\ref{rmax_eq_eta} to be analytically tractable, we need to 
simplify the expression for $\kappa_0$ (Eq.~\ref{kappa0_eq}). For an isothermal 
ellipsoid, the magnitude of the deflection (which is proportional to $\kappa_0 
\theta$) is independent of radius; while this is not strictly true for 
$\alpha\neq 1$, $\kappa_0\theta$ will nevertheless vary slowly with radius 
provided the slope $\alpha$ is not too extreme.  Therefore, we expand $\kappa_0 
\theta$ to first order in $\theta$ around the position of the subhalo 
$\theta_s$, which gives (after dividing by $\theta$),

\begin{equation}
\kappa_0(\boldsymbol{\theta_s} + \boldsymbol{r}) \approx \kappa_0(\boldsymbol{\theta_s})\left[1 - \alpha\frac{r}{\theta_s+r}\right].
\end{equation}

Plugging this into Eq.~\ref{rmax_eq_eta} results in the following equation, 
which can now be solved analytically for an assumed subhalo density profile:

\begin{equation}
\eta - (1+g)\kappa_0(\boldsymbol{\theta_s}) + \frac{\alpha 
(1+g)\kappa_0(\boldsymbol{\theta_s}) r}{\theta_s + r} = \bar\kappa_s(r).
\label{general_rmax_eq}
\end{equation}

\subsection{B2. Defining the effective subhalo lensing mass}

Before assuming a density profile for the subhalo, we note that in the absence 
of the subhalo ($\bar\kappa_s=0$), Eq.~\ref{rmax_eq_eta} tells us the position 
of the (unperturbed) critical curve $\theta_c$ along the direction $\phi$ is 
given by $\eta = (1+g)\kappa_0(\theta_c)$. Since the subhalo will be close to 
the unperturbed critical curve, we can expand 
$(1+g)\kappa_0(\boldsymbol\theta_s)$ around $\theta_c$, which to second order 
is

\begin{equation}
(1+g)\kappa_0(\boldsymbol\theta_s)\approx \eta\left[1 - \alpha\frac{\Delta\theta_s}{\theta_c} + \frac{\alpha(\alpha+1)}{2}\left(\frac{\Delta\theta_s}{\theta_c}\right)^2\right]
\label{kappa0_theta_expansion}
\end{equation}
where $\Delta\theta_s = \theta_s - \theta_c$. For the moment, we will keep only first-order terms in $\Delta\theta_s/\theta_c$ and $r/\theta_s$. Combining Eqs.~\ref{kappa0_theta_expansion} and \ref{general_rmax_eq} and keeping first-order terms, we find

\begin{equation}
\frac{\Delta\theta_s}{\theta_c} + \frac{r}{\theta_s} \approx \frac{\bar\kappa_s(r)}{\alpha\eta}.
\end{equation}
This result is very important: it implies that if a lens model reproduces the correct subhalo perturbation radius $r = \delta\theta_c$ (which is the solution to the above equation), and if the best-fit position of the subhalo is accurate, the quantity $\bar\kappa_s(r)/\alpha\eta$ will be well-reproduced regardless of the mass distribution of the subhalo, and even if the best-fit $\alpha$ and external shear $\Gamma_{ext}$ differ from their actual values. Since the average kappa is proportional to the projected subhalo mass $m_{sub}(r)$ enclosed within a given radius, we can say that

\begin{equation}
\left(\frac{m_{sub}(\delta\theta_c)}{\alpha\eta}\right)_{fit} \approx \left(\frac{m_{sub}(\delta\theta_c)}{\alpha\eta}\right)_{true}.
\label{invariant_mass_eq}
\end{equation}

In general, after fitting a lens, it is impossible to know to what extent the 
true value for the external shear differs from the best-fit value. However, 
since $\eta \approx 1$ regardless, in practice this effects 
Eq.~\ref{invariant_mass_eq} only slightly, typically not more than a few 
percent.  The effect of $\alpha$ can be more dramatic, but since most 
gravitational lens galaxies are well-fit by an isothermal profile, in practice 
one can surmise that $\alpha_{true} \approx 1$ and find the mass enclosed 
within the subhalo under this assumption. Thus, we define the \emph{effective 
subhalo lensing mass} $\tilde m_{sub}$ to be the projected mass enclosed within 
the subhalo perturbation radius, divided by $\alpha$; in other words,

\begin{equation}
\tilde m_{sub} \equiv m_{sub}(\delta\theta_c)/\alpha.
\label{}
\end{equation}

In Section \ref{sec:results} we show that this mass estimator is robust even if 
the assumed density profile and/or tidal radius of the subhalo are inaccurate.

\subsection{B3. Solution for the subhalo perturbation radius under the 
assumption of a Pseudo-Jaffe subhalo}

Now that we have established the usefulness of the subhalo perturbation radius 
$\delta\theta_c$, we will now assume the subhalo has a Pseudo-Jaffe density 
profile with a tidal radius $r_t$, for which we have

\begin{equation}
\bar\kappa_s(r) = \frac{b_s}{r^2}\left(r + r_t - \sqrt{r_t^2+r^2}\right).
\end{equation}
In order to render the equation analytically tractable, we Taylor expand the 
deflection $r\bar\kappa$ to first order. Since a rough approximation for 
$\delta\theta_c$ is given by $r \approx \delta\theta_{c} \approx \sqrt{b b_s}$, 
we expand around this point. (The approximate solution we expand around can be 
improved upon, e.g. by taking the highest-order term in 
Eq.~\ref{rmax_approx_eq2}, which typically leads to an extra 1\% percent 
accuracy in $\delta\theta_c$; however, in the interest of simplicity we choose 
$\sqrt{b b_s}$ instead.) The result is
\begin{equation}
\bar\kappa_s ~ \approx ~ \epsilon\frac{b_s}{r} - \xi\sqrt{b_s/b}
\label{kapavg_subhalo_expansion}
\end{equation}
where
\begin{equation}
\epsilon = 1 + 2\beta - 2\left(1+\beta^2\right)^{\frac{1}{2}} + \left(1+\beta^2\right)^{-\frac{1}{2}},
\label{epsilon_eq}
\end{equation}
\begin{equation}
\xi = \beta - \left(1+\beta^2\right)^{\frac{1}{2}} + \left(1+\beta^2\right)^{-\frac{1}{2}},
\label{xi_eq}
\end{equation}
and $\beta$ is defined as the ratio of the subhalo's distance $r_0$ to the 
center of the host galaxy over the Einstein radius,
\begin{equation}
\beta \equiv \frac{r_0}{b} = \frac{r_t}{\sqrt{bb_s}}.
\label{beta_eq}
\end{equation}
The latter equality is given by the formula for the Jacobi radius of an 
isothermal subhalo. For many subhalos, we can expect that $\beta \gg 1$, and so 
a reasonable approximation in many cases may be obtained by taking $\beta 
\rightarrow \infty$, which gives $\epsilon \approx 1$, $\xi \approx 0$.

As we discussed in Section \ref{sec:tidal_radius}, it has been customary to 
assume that $r_0 \approx b$ (so that $\beta \approx 1$), based on the fact that 
the projected distance of the subhalo to the host galaxy's center is comparable 
to the Einstein radius.  We emphasize that this assumption is unlikely to hold, 
because the Einstein radius is typically in the range of 5-10 kpc and a subhalo 
would be unlikely to survive the severe tidal stripping that would result from 
being so close to the host galaxy.  However, for the sake of comparison to 
previous work, we will consider the results using this assumption, in which 
case $r_t \approx \sqrt{bb_s}$ and we have
\begin{equation}
\epsilon = 3\left(1 - \frac{1}{\sqrt{2}}\right), ~ ~ \xi = \frac{\epsilon}{3} ~ ~ ~ ~ ~ (\textrm{for} ~ r_0 = \textrm{Einstein radius}).
\label{eq_epsilon_eta}
\end{equation}

Substituting Eqs.~\ref{kapavg_subhalo_expansion} and 
\ref{kappa0_theta_expansion} into Eq.~\ref{general_rmax_eq} and solving the 
resulting quadratic equation in $r$, we find the following solution,

\begin{equation}
\delta\theta_{c} \approx \frac{1}{\mu}\left(\sqrt{b_s\theta_s\epsilon\mu + \left(\zeta/2\right)^2} + \zeta/2\right),
\label{rmax_general}
\end{equation}
where
\begin{equation}
\mu = \alpha\eta\left[1 + \frac{\xi}{\alpha\eta}\sqrt{\frac{b_s}{b}} + (1-\alpha)\psi\right]
\label{mu_eq}
\end{equation}
\begin{equation}
\zeta = b_s\epsilon - \xi\theta_s\sqrt{\frac{b_s}{b}} - \alpha\eta\theta_s\psi,
\label{zeta_eq}
\end{equation}
\begin{equation}
\psi = \frac{\Delta\theta_s}{\theta_c}\left(1 - \frac{1}{2}(\alpha+1)\frac{\Delta\theta_s}{\theta_c}\right).
\label{psi_eq}
\end{equation}

To use these formulae, we need to calculate the position of the unperturbed 
critical curve $\theta_c$ along the direction of the subhalo. This can be found 
numerically using gravitational lensing software, in which case there is no 
need to calculate $g$ at all since it does not appear explicitly in 
Eqs.~\ref{rmax_general}-\ref{psi_eq}.  Alternatively, $\theta_c$ can be found 
analytically by solving the equation $\eta = 
(1+g)\kappa_0(\boldsymbol\theta_s)$. By substituting Eq.~\ref{kappa0_eq}, one 
can show that

\begin{equation}
\theta_c = b\sigma\left[\frac{1}{2\eta}(1+g)(2-\alpha)\right]^{1/\alpha}.
\end{equation}

We have tested Eq.~\ref{rmax_general} in many different scenarios and find that 
it is accurate to within 2\%, provided the external shear is not too large 
($\Gamma_{ext} \lesssim 0.2$). In order to test our formula on the simulated 
lenses we modeled in Section \ref{sec:results}, in 
Table~\ref{tab:dtheta_c_comparison} we list the numerical solution (first 
column) for $\delta\theta_c$ along with the values given by 
Eq.~\ref{rmax_general} (second column) for the best-fit model in each 
simulation.

\begin{table}
\centering
\begin{tabular}{|l|c|c|c|}
\hline
 & $\delta\theta_{c}$ (exact) & $\delta\theta_{c}$ (Eq.~\ref{rmax_general})& $\delta\theta_{c}$ (approx, Eq.~\ref{rmax_approx_eq2})\\
\hline
Sim. 1 & 0.149'' & 0.147'' & 0.185'' \\
Sim. 2 & 0.155'' & 0.151'' & 0.188'' \\
Sim. 3 & 0.175'' & 0.171'' & 0.202'' \\
Sim. 4 & 0.161'' & 0.157'' & 0.193'' \\
\hline
\end{tabular}
\caption{Comparison between different solutions for the subhalo perturbation 
radius $\delta\theta_c$ of the best-fit model in each simulated lens scenario 
(see Table \ref{tab:sims} for a description of each simulation). We list the 
numerical solution (first column), analytic solution Eq.~\ref{rmax_general} 
(second column) and approximate analytic solution Eq.~\ref{rmax_approx_eq2} 
(third column).}
\label{tab:dtheta_c_comparison}
\vspace{10pt}
\end{table}

Since Eq.~\ref{rmax_general} is still cumbersome to use for non-lens modelers, 
it is useful to have a simpler approximate expression.  If we approximate 
$\Delta\theta_s \approx 0$ (and hence $\psi \approx 0$) in the above equations 
and expand to first order in $\sqrt{b_s/b}$, we find

\begin{equation}
\delta\theta_{c} \approx \sqrt{\frac{\epsilon\theta_s b_s}{\alpha\eta}}\left(1 
- \frac{x}{2} + \frac{x^2}{8}\right) + \frac{\epsilon b_s}{2\alpha\eta}(1-x)
\label{rmax_approx_app}
\end{equation}
where
\begin{equation}
x = \xi\sqrt{\frac{\theta_s}{b\epsilon\alpha\eta}}.
\label{x_eq_app}
\end{equation}

One may set $\eta \approx 1$ in 
Eqs.~\ref{rmax_approx_app} and \ref{x_eq_app} with at most a few percent loss 
of accuracy, so we reproduce this formula in Section 
\ref{sec:subhalo_perturbation} with $\eta$ set to one (Eq.~\ref{rmax_approx_eq2}).  
If $\Delta \theta_s$ is indeed zero, we find this formula (with $\eta$ set to 
1) is accurate to within $\sim$1\% provided the external shear is not too large 
($\Gamma_{ext} \lesssim 0.1$). The further away the subhalo is from the 
critical curve, the worse this approximation becomes---generally it will give a 
value for $\delta\theta_c$ that is too large. However, in the simulated lenses 
we examine in Section \ref{sec:results}, the optimal radius where the effective 
subhalo lensing mass is invariant is a bit larger than the actual value for 
$\delta\theta_c$, and this may be true in general. One can see from Figure 
\ref{fig:mprofiles} that Eq.~\ref{rmax_approx_app} actually works better than 
using the actual $\delta\theta_c$, even though it is greater by about 17-25\% 
depending on the simulation. The values given by this approximation are listed 
in the third column of Table~\ref{tab:dtheta_c_comparison} for the best-fit 
model in each simulation considered in Section \ref{sec:results}.

Incidentally, one may verify from Eq.~\ref{rmax_approx_app} that if the 
log-slope of the host galaxy $\alpha$ and external shear $\eta$ are varied, 
then to get the same $\delta\theta_{c}$ (to highest order) the Einstein radius 
of the subhalo $b_s$ must be scaled by a factor $1/\alpha\eta$; in other words, 
to keep the same perturbation scale, $b_s/\alpha\eta$ must be invariant.  This 
is another manifestation of the fact that 
$\bar\kappa_s(\delta\theta_{c})/\alpha\eta$ is invariant, as follows from 
Eq.~\ref{invariant_mass_eq}.

\end{document}